\documentclass[12pt,draftclsnofoot,onecolumn]{IEEEtran}

\usepackage{amsmath,amssymb,dsfont,stfloats,amsthm,color,subfigure,cite}
\usepackage{hyperref}
\usepackage[pdftex]{graphicx}

\newtheorem{thm}{Theorem}

\newtheorem{lem}{Lemma}
\newtheorem{corol}{Corollary}

\newtheorem{defi}{Definition}
\newtheorem{rem}{Remark}

\theoremstyle{definition}

\theoremstyle{remark}

\theoremstyle{ex}
\newtheorem{ex}{Example}

\def\argmax{\mathop{\rm argmax}}

\def\pa{{\pmb a}}\def\pd{{\pmb d}}\def\ph{{\pmb h}}\def\px{{\pmb x}}\def\py{{\pmb y}}\def\pz{{\pmb z}}

\def\pA{{\pmb A}}\def\pB{{\pmb B}}\def\pH{{\pmb H}}\def\pI{{\pmb I}}\def\pP{{\pmb P}}\def\pR{{\pmb R}}\def\pS{{\pmb S}}\def\pU{{\pmb U}}\def\pW{{\pmb W}}\def\pX{{\pmb X}}\def\pY{{\pmb Y}}\def\pZ{{\pmb Z}}\def\p0{{\pmb 0}}

\def\ct{\mathsf{H}}

\def\snr{{\small\textsf{SNR}}}

\long\def\comment#1{}

\newfont{\bbb}{msbm10 scaled 700}

\newfont{\bbc}{msbm10 scaled 1100}

\newcommand{\wv}{{\pmb w}}

\newcommand{\Am}{{\pmb A}}
\newcommand{\Bm}{{\pmb B}}

\newcommand{\Hm}{{\pmb H}}
\newcommand{\Id}{{\pmb I}}

\newcommand{\Sm}{{\pmb S}}

\newcommand{\Vm}{{\pmb V}}
\newcommand{\Wm}{{\pmb W}}
\newcommand{\Xm}{{\pmb X}}

\newcommand{\Ac}{{\cal A}}

\newcommand{\Cc}{{\cal C}}

\newcommand{\Lc}{{\cal L}}

\newcommand{\Sc}{{\cal S}}

\newcommand{\Lambdam}{\hbox{\boldmath$\Lambda$}}

\newcommand{\Sigmam}{\hbox{\boldmath$\Sigma$}}

\newcommand{\diag}{{\hbox{diag}}}

\newcommand{\trace}{{\hbox{tr}}}

\IEEEoverridecommandlockouts

\begin{document}

\title{On the Role of Transmit Correlation Diversity\\ in Multiuser MIMO Systems}

\author{Junyoung Nam, Giuseppe Caire, Young-Jo Ko, and Jeongseok Ha
\thanks{This work was supported by the ICT R\&D program of MSIP/IITP [14-000-04-001]. The material in this paper was presented in part  at the IEEE International Symposium on Information Theory (ISIT),  Jun./Jul. 2014.}
\thanks{J. Nam is with the Wireless Communications Division, Electronics and Telecommunications Research Institute (ETRI), Daejeon, Korea, and he is also with the Department of Electrical Engineering, Korea Advanced Institute of Science and Technology (KAIST), Daejeon, Korea (e-mail: jynam@etri.re.kr). } 
\thanks{G. Caire is with the Communication Systems Department, Faculty IV, Technical University of Berlin,
Germany (e-mail: caire@tu-berlin.de).}
\thanks{Y. -J. Ko is with the Wireless Communications Division, ETRI, Daejeon, Korea (e-mail: koyj@etri.re.kr).}
\thanks{J. Ha is with the Department of Electrical Engineering, KAIST, Daejeon, Korea (e-mail: jsha@kaist.edu).}
}

\maketitle

\begin{abstract}
Correlation across transmit antennas, in multiple antenna systems (MIMO), 
has been studied in various scenarios and has been shown to be detrimental or provide benefits depending on the 
particular system and underlying assumptions. In this paper, we investigate the effect of transmit correlation on the capacity of the 
Gaussian MIMO broadcast channel (BC), with a particular interest in the large-scale array (or massive MIMO) regime. 
To this end, we introduce a new type of diversity, referred to as \emph{transmit correlation diversity}, which captures the fact that the channel vectors of different users 
may have different, and often nearly mutually orthogonal, large-scale channel eigen-directions. 
In particular, when taking the cost of downlink training properly into account, transmit correlation diversity can yield significant capacity gains in all regimes of interest. 
Our analysis shows that the system multiplexing gain can be increased by a factor up to $\lfloor{M}/{r}\rfloor$, where $M$ is the number of antennas and $r\le M$ is the common rank of 
the users transmit correlation matrices, with respect to  standard schemes that are agnostic of the transmit correlation and treat the channels as if they were isotropically distributed. 
Thus, this new form of diversity reveals itself as a valuable ``new resource" in multiuser communications.
\end{abstract}

\begin{keywords}
Broadcast channel, random matrix theory, transmit antenna correlation, large-scale (massive) MIMO.
\end{keywords}

\section{Introduction}
\label{sec:intro}

Traditionally, channel fading had been considered as a harmful source to combat with transmit or receive diversity. However, since the seminal work of \cite{Fos96,Tel99}, 
independent fading in multiple-antenna (MIMO) systems has been shown to provide large capacity gains, thanks to the fact that the number of degrees of freedom of 
such MIMO channels grows with the minimum of the transmit and receiving antennas. 
Focusing on downlink communication, we consider the case where the transmitter (base station) is equipped with a number of antennas, and the receivers (users) are equipped with
a single antenna.\footnote{In practice, a user device may be equipped with multiple antennas, which are then combined in order to form a beamforming pattern, i.e., 
a directional antenna, to achieved beamforming gain and/or inter-cell interference rejection. What matters for the analysis in this paper is that each user receives a single data stream, 
i.e., even though the device is equipped with multiple antennas, their output is combined and demodulated as a single stream.}
In a typical system geometry where the base station array is mounted on a tower or on a relatively tall building, and the propagation between the base station and the users occurs along 
clusters of scatterers that are seen from the base station on a narrow angular spread, the coefficients of the channel vector describing the propagation between the base station array and 
a given user are correlated Gaussian random variables. In contrast, the channel vectors of different users, which are physically separated by multiples of the wavelength,\footnote{For example, 
for a typical carrier frequency between 2GHz and 5GHz, the channel wavelength is between 15 and 6cm.} are mutually statistically independent. 
Spatially correlated MIMO channels have been well characterized for a variety of transmit correlation models \cite{Shi00,Chu02,Ver05,Tul05}. 
Traditionally, transmit correlation has been considered to be a detrimental source, thereby incurring power loss at high signal-to-noise ratio (SNR) (e.g., \cite{Loz05}). 
Some exceptions where transmit correlation helps capacity are the case of low SNR \cite{Ver05,Tul05}, where the capacity-achieving input covariance is non-isotropic, and the case where
channel state information (CSI) is not available at all \cite{Jaf05}, for which knowing the statistics of the channel may effectively help.  
The impact of transmit correlation on the ergodic capacity is much less known in the multiuser context, albeit the capacity region of the Gaussian MIMO BC with perfect CSI at both transmitter 
and receivers is fully understood \cite{Wei06} irrespectively of the channel statistics. The work of \cite{Aln09} extended the sum-rate scaling result of \cite{Sha05} to the special case where 
all users have a common channel covariance matrix, and concluded that transmit correlation has a detrimental impact on the sum capacity of multiuser MIMO (MU-MIMO) systems.

A different line of works has pointed out that transmit correlation can be in fact advantageous for MU-MIMO systems, with respect to aspects such as CSI feedback overhead, 
scheduling, and codebook design \cite{Ham08, Tri08, Cle08,Lee09}). The key observation is that, in a multiuser environment, there exist \emph{diverse} transmit correlations across 
multiple users. Basically, different transmit correlations indicate different ``large-scale" (or long-term) preferential directions of the user channels, depending on the geometry of local scatterers 
and transmit antenna spacing. Therefore, the diversity of transmit correlations can be leveraged in the multiuser communication framework. 
In order to fully exploit such effect, the authors in \cite{Nam12} introduced an optimistic condition by imposing a useful structure on channel covariances of users, referred to as \emph{tall unitary structure}, for the MU-MIMO downlink. 
The resulting joint spatial division and multiplexing (JSDM) precoding strategy was extensively studied to show that the tall unitary condition holds asymptotically 
at least for the case of uniform linear array (ULA) with large number of antennas, 
whenever the user groups have scattering with disjoint angular support \cite{Adh13}. Further work along this line of research has considered: 1) the system capacity in the large system regime 
(both the number of antennas and the number of users grow to infinity at a fixed ratio) \cite{Nam13a}; 2) schemes for opportunistic beamforming with probabilistic scheduling \cite{Nam14}; 
3) the suitability of the JSDM framework for millimeter wave (mm-Wave) channels \cite{Adh13b}; 4) the elimination of pilot contamination in multi-cell TDD systems where different cells share the same 
set of uplink pilot signals \cite{Yin13};  5) how to design coordinated composite beamforming schemes where the knowledge of the transmit correlation can be used 
to improve the performance of multi-cell MU-MIMO networks \cite{Liu14,adhikary2014massive}.

In this paper, we coin the term \emph{transmit correlation diversity} for the new type of diversity under the ideal structure, where users are partitioned into groups such that all users in the same group have the same channel covariance and the eigenspaces spanned by the channel covariances of groups are mutual orthogonal (or {linearly independent} for a weaker condition).  Assuming this unitary condition and the symmetric case where all group covariances are of rank $r\le M$, where $M$ denotes the number of base station antennas, 
the number of degrees of transmit correlation diversity can be expressed as $G = \lfloor{M}/{r}\rfloor$. 
While previous related works paid great attention to lower bounds on the performance gain achieved by exploiting transmit correlation, e.g., by focusing on the achievable sum-rate 
analysis of JSDM for large-scale MU-MIMO systems in \cite{Nam12,Adh13,Nam13a,Nam14,Adh13b},  in this work we rather focus on information-theoretic upper bounds 
in order to provide some new insights into the role of transmit correlation diversity with respect to the capacity of MIMO BCs. 
Specifically, we wish to answer to the following questions: In which regimes of the system parameters including $M$, 
the number of users $K$, $G$, and $T_c$, where $T_c$ is the coherence time interval, 
can transmit correlation diversity be beneficial to the capacity? What are the upper bounds on its potential gain in the {regimes of interest}, compared to the capacity of the independent and identically 
distributed (i.i.d.) Rayleigh fading MIMO BC? 

Assuming perfect CSI, we first study the impact of transmit correlation diversity on the power gain (the parallel shift of capacity versus SNR curves, also known as power offset) 
of MIMO BC. The authors in \cite{Nam13a,Nam14} characterized the asymptotic capacity behavior in the large number of users regime. In sharp contrast to \cite{Aln09}, it turned out that transmit correlation diversity can achieve a sum-rate gain of up to $M\log G$ (i.e., power gain of $3\log G$ dB). However, it was not fully understood why we could do better than the i.i.d. Rayleigh fading case in this regime. 
In addition, it was not known whether transmit correlation diversity can achieve capacity gains in other regimes of the system parameters, compared to the independent fading case. 
To this end, we need to investigate the impact of transmit correlation on the power gain of MIMO BC in various regimes. It turns out that transmit correlation diversity may be 
even harmful, especially when $M$ is larger than $K$, under the idealized assumption of perfect CSI at the base station.

Taking the cost of downlink training into consideration rather than assuming perfect CSI, the well-known limit on the sum rate of the \emph{i.i.d. Rayleigh block-fading} MIMO channel immediately 
provide a cut-set upper bound, following from the work of Zheng and Tse \cite[Sec. V]{Zhe02} (see also \cite{Has03}), as already noticed in \cite{Huh12,Adh13}. 
Namely, the high-SNR capacity of the resulting pilot-aided systems is limited by 
$  M_\text{iid}^*(1-M_\text{iid}^*/T_c)\log \snr+O(1),$ where $M_\text{iid}^* = \min\{M,K,\lfloor T_c/2\rfloor\}.$
For typical cellular downlink systems with $M$ small, where $\min\{M,K\} \ll T_c$, the factor $T_c/2$ does not significantly affect the system performance. 
However, in the large-scale array regime with $M > T_c$, to which a great deal of attention has been paid since \cite{Mar10}, this factor is shown to have a critical impact on the system 
performance. To be specific, no matter how large both $M$ and $K$ are, multiplexing gain is fundamentally \emph{saturated} by $T_c/4$ when $T_c/2 \le \min(M,K)$. Interestingly, this limit is obtained by letting only $T_c/2$ users send uplink pilot signals on the first half of the coherence block, i.e., for
$T_c/2$ dimensions, and using the remaining $T_c/2$ dimensions to serve these users by spatial multiplexing. Thanks to the TDD reciprocity, $M$ can be made arbitrarily large, thus
going to the regime of massive MIMO. Notice that this scheme that uses half of the coherence block for uplink training and the other half for downlink data transmission is
precisely what was provided as an LTE-motivated example in \cite{Mar10}. As a result of this upper bound, when both $M$ and $K$ are large, 
the coherence time $T_c$ becomes the system dimensionality bottleneck, which fundamentally limits the system multiplexing gain. 
Note that the above upper bound in \cite{Zhe02} holds only for isotropically distributed channels. 
As a consequence, any scheme based on uplink training and TDD reciprocity that is agnostic of the channel statistics, i.e., which treats the channels as if they were 
independent isotropically distributed random vectors, must also obey to such bound. In contrast,  in this work we shall show that this is not necessarily the case in spatially correlated 
fading channels, for which it is possible to take explicit advantage of the knowledge of the channel statistics (i.e., of the covariance matrix) in order to 
break the above dimension bottleneck. 
In particular, we find that the multiplexing gain can continue to grow as $M$ and $K$ increase, 
provided that the degrees of transmit correlations diversity are sufficiently large.  Therefore, transmit correlation diversity is indeed beneficial to significantly increase the multiplexing gain of MU-MIMO 
systems, as well as the power gain in some regimes. By taking the CSI estimation (downlink training) overhead into account, 
we show that transmit correlation diversity is beneficial in all regimes of the system parameters, apart from the case where $\min\{M,K\}$ is too small. 

The remainder of this paper is organized as follows. Section \ref{sec:Pre} describes the MU-MIMO downlink system model of interest and briefly reviews a key result of JSDM with the notion of transmit correlation diversity. In Section \ref{sec:IT}, we study the impact of transmit correlation to the power gain of MIMO BCs in several regimes of system parameters, assuming perfect CSI. Section \ref{sec:FL} investigates the impact of transmit correlation to the multiplexing gain of pilot-aided MU-MIMO systems, where the cost of downlink training is considered. 
We conclude this work in Section \ref{sec:CR}.

{\em Notation}: 
$\pA^\ct$ and $\lambda_i(\pA)$ denote the Hermitian transpose and the $i$th eigenvalue (in descending order) of matrix $\pA$. $\trace(\pA)$ and $|\pA|$ denote the trace and the determinant of a square matrix $\pA$. $\pI_n$ denotes the $n \times n$ identity matrix.  $\|\pa\|$ denotes the $\ell_2$ norm of vector $\pa$. We also use  $\px \sim \mathcal{CN}(\p0;\Sigmam)$ to indicates that $\px$ is a zero-mean complex circularly-symmetric Gaussian vector with covariance $\Sigmam$. $\mathbb{Z^+}$ denotes the set of positive integers. The base of the logarithms used in this work is $2$. Finally, let $\Cc^\text{sum}(\Ac)$, where $\Ac$ is a subset of $\snr$, $M$, $K$, $r$, and $G$, denote the asymptotic sum capacity of MIMO BC when system parameters in the set $\Ac$ are sufficiently large.

\section{Preliminaries}
\label{sec:Pre}

\subsection{System Model}
\label{sec:SD}

Consider a MIMO BC (downlink) with $M$ transmit antennas and $K$ users equipped with a single antenna
each. For spatial correlation between transmit antennas, we use the well-known Kronecker model \cite{Shi00,Chu02} (or separable correlation model) $\pH={\pR}^{\frac{1}{2}} \Wm,$ where there is no receive correlation due to single-antenna users, the elements of $\Wm$ are i.i.d. $\sim \mathcal{CN}(0,1)$, and $\pR$ denote the deterministic transmit correlation matrices, respectively, assuming the wide-sense stationarity of the channels. The random matrix $\pH$ follows the frequency-flat block-fading model for which it remains constant during the coherence time interval of $T_c$ but changes independently every interval. Most of our results in this paper remain valid in the more general unitary-independent-unitary model (for which see \cite{Tul05}), since the elements of $\pW$ are allowed to be independent \emph{nonidentically} distributed to apply some well-known results of random matrix theory to be used in this paper. 

In this paper, we let $\pR$ normalized as $\trace(\pR)=M$ without loss of generality for all users. 
By using the Karhunen-Loeve transform, the channel vector of a user can be expressed as
\begin{align} \label{eq:SM-2}
   \ph={\pU}{\boldsymbol \Lambda}^{\frac{1}{2}} \wv
\end{align}
where 
${\boldsymbol \Lambda}$ is an $r\times r$ diagonal matrix whose elements are the non-zero eigenvalues
of $\pR$, $\pU \in \mathbb{C}^{M\times r}$ is a tall unitary matrix whose columns are the eigenvectors of $\pR$ corresponding to the non-zero eigenvalues, i.e., $\pR=\pU\Lambdam\pU^\ct$, and $\wv \in \mathbb{C}^{r \times 1} \sim\mathcal{CN} (\p0, \pI)$. 

Let $\pH$ denote the $M\times K$ system channel matrix given by
stacking the $K$ users channel vectors $\ph$ by columns. The signal vector received by the users is given by
\begin{align} \label{eq:SM-3}
   \py=\pH^\mathsf{H}\Vm\pd +\pz =\pH^\mathsf{H}\px +\pz
\end{align}
where $\Vm$ is the $M\times s$ precoding matrix with $s$ the rank of the input covariance
$\boldsymbol{\Sigma}=\mathbb{E}[\px\px^\ct]$ (i,e., the total number of independent data streams),
$\pd$ is the $s$-dimensional transmitted data symbol vector such that the transmit signal vector is given by $\px=\Vm \pd$, and $\pz  \sim\mathcal{CN} (\p0, \pI)$ is the Gaussian noise
at the receivers. The system has the total power constraint such that $\trace(\Sigmam)\le P$, where $P$ implies the total transmit SNR.

\subsection{Summary of JSDM}

We briefly review the JSDM strategy \cite{Nam12} that was originally introduced to reduce the cost for downlink training and CSI feedback in FDD large-scale MIMO systems by exploiting the fact that some users have similar transmit correlation matrices.
The idea is to group together users with similar transmit correlations and then separate the different groups by a pre-beamforming matrix which is calculated only as a function of the channel second-order 
statistics, and does not depend on the instantaneous CSI. This creates a sort of {spatial division}, that exports the fact that the ``long-term" preferential direction of the channel vectors of users belonging to different groups are nearly mutually orthogonal.  In general, we have multiple sets of quasi-orthogonal groups, which we call \emph{classes}. 
Each class $t$ is served separately over an orthogonal transmission resource (i.e., a time-frequency slot) and may have a different number of groups, denoted by $G_t$. 
Therefore, we partition the entire user set, $\mathcal{K}=\{1,2,\cdots,K\}$, into $T$ non-overlapping subsets (classes).

As anticipated before, the JSDM precoder $\Vm = \pB \pP$ is formed by two-stages 
$\pB$ and $\pP$.  The first stage, referred to as pre-beamforming, consists of a matrix
$\pB$ of dimensions $M \times b$, where $b\le M$ is an intermediate dimension whose optimization is discussed in \cite{Adh13}. 
The pre-beamforming matrix depends only on the channel second-order statistics, i.e., on the channel covariance matrices of the different groups served simultaneously
by spatial multiplexing (i.e., belonging to the same class). Since $\pB$ depends only on the second-order statistics, which is very slowly time-varying,\footnote{Strictly speaking, for the classical Wide-Sense Stationary (WSS) fading channel model, the second-order statistics is time-invariant and therefore it can be estimated by time-averaging at negligible overhead cost. As a matter of fact, 
due to non-stationary effects such as large user motion, and users joining and leaving the system, the WSS assumption holds only ``locally'' on a time scale which is anyway 
orders of magnitude slower than the channel coherence time $T_c$. Therefore, while the estimation and tracking of the channel covariance matrix is an interesting topic in itself, it is safe to assume here that 
this is known at no significant overhead dimensional cost.} it can be considered as perfectly known by the base station. 
The second stage consists of a MU-MIMO precoding matrix $\pP$, of dimension $b\times s$, determined as a function of the 
instantaneous realization of the projected effective channel $\pH^\mathsf{H}\pB$.
We divide $b$ and $s$ such that $b=\sum_g b_g$ and $s = \sum_g s_g$, where $b_g \geq s_g$ for all $g = 1,\ldots, G$,
and denote by $\pB_g$ the $M \times b_g$ pre-beamforming matrix of group $g$.
Thanks to the above user partitioning, we can consider estimating only the $G$ diagonal blocks 
\begin{align} \label{eq:SM-10}
   \textsf{\pH}_g \triangleq \Bm_g^\mathsf{H} \Hm_g, \ g=1,\cdots,G
\end{align} 
where $\pH_g$ is the aggregate channel matrix given by stacking the channel vectors of users in group $g$,  and each group is independently processed by treating signals 
of other groups as interference.  In this case, the MU-MIMO precoding stage takes on the block-diagonal form
$\pP=\mathrm{diag}(\pP_1,\cdots,\pP_G)$, where $\pP_g\in \mathbb{C}^{b_g \times s_g}$, yielding the vector broadcast plus interference Gaussian 
channel 
\begin{equation} 
\py_g = {\pH}_g^\mathsf{H}\pB_g \pP_g\pd_g +  \sum_{h \neq g}  {\pH}_g^\mathsf{H}\pB_{h} \pP_{h}\pd_{h}   +  \pz_g, 
\end{equation} 
for $g = 1,\ldots, G$. 

\subsection{Transmit Correlation Diversity}
\label{sec:TCD}

\begin{figure}
\center \includegraphics[scale=.9]{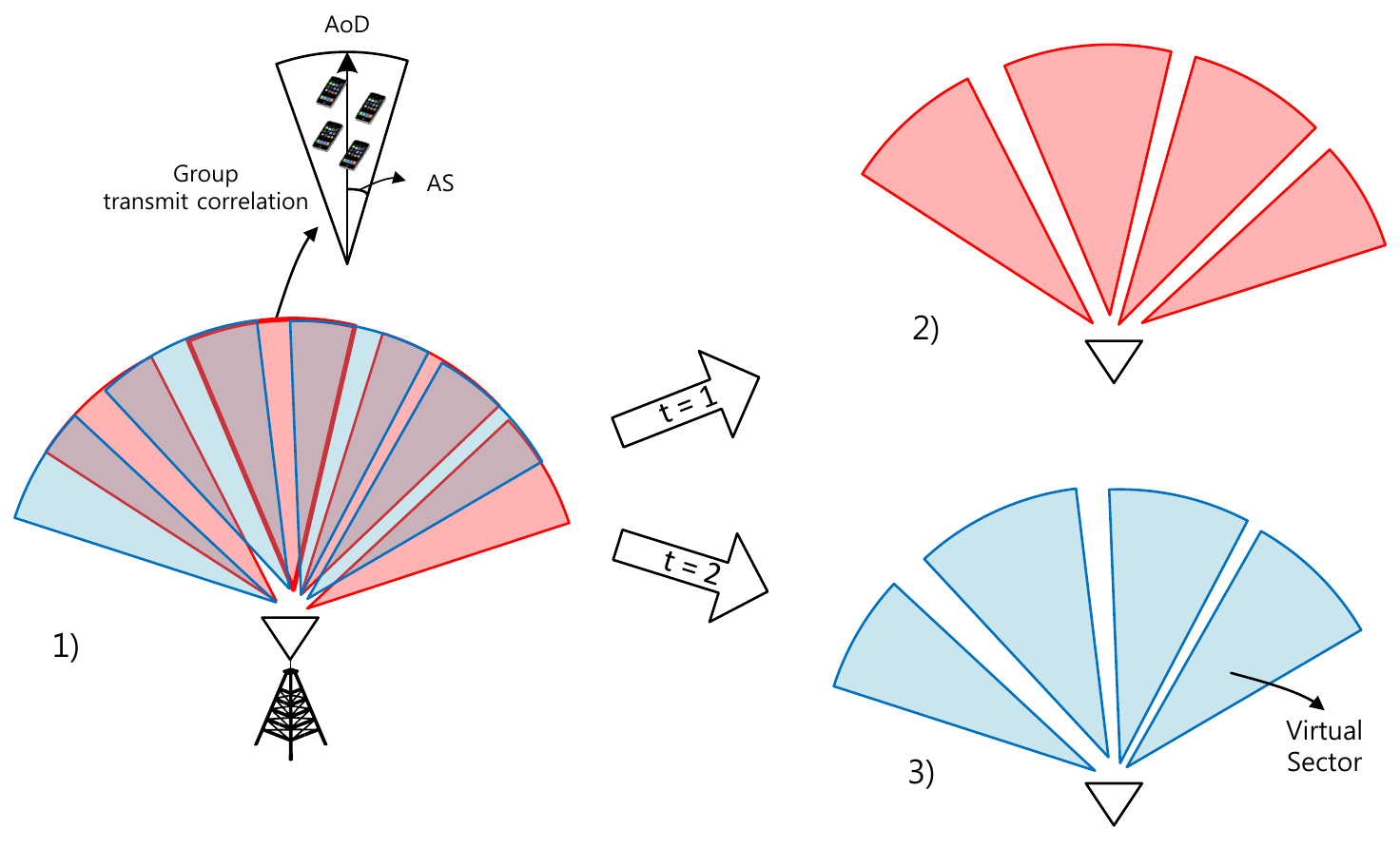} 
\caption{Illustration of virtual sectorization exploiting transmit correlation diversity with $T=2$ and $G=4$.}  \label{fig-0}
\end{figure}

We introduce the notion of transmit correlation diversity to better understand the key idea of JSDM. Fig. \ref{fig-0} depicts a simple example which explains the virtual sectorization enabled by exploiting diverse transmit correlations in a three-sector BS with $D=1/2$ in the following steps.  

\begin{enumerate}
\item In the beginning, angular regions (pie slices drawn by AoD\footnote{AoD and angle of arrival (AoA) are generally different in FDD. As AoD is more precise at the transmitter side, we prefer the terminology AoD to AoA.} 
 and AS at the BS) roughly representing long-term eigenspaces are overlapped, i.e., user groups are interfering with each other and there is no noticeable structure. 
\item Put together the red angular regions into class $t=1$ and separate them by multiple pre-beamforming along their respective eigenspaces. By doing so, each group can be viewed as a virtual sector.
\item Do the same thing on the blue regions for class $t=2$.
\item Multiple users within each group (i.e., virtual sector) can be simultaneously served by the second-stage MU-MIMO precoding.
\end{enumerate}

Given the geometric intuition provided by the clustered scattering correlation model (e.g., one-ring model \cite{Shi00}), we define transmit correlation diversity of a multiuser system as follows:

\begin{defi}[Transmit Correlation Diversity] \label{def-3}
A multiuser MIMO downlink system after user partitioning is said to have $G$ degrees of transmit correlation diversity, if the eigenspaces of $G_t$ groups in class $t$ are mutually linearly independent for all classes and $G=\frac{1}{T}\sum_{t=1}^T G_t$.  
\end{defi}

Notice that the number of groups per class, $G_t$, must be an integer by definition.
Although transmit correlation diversity is formally defined conditioned on the linear independence between the group eigenspaces, the effect of transmit correlation diversity does not necessarily require the condition, as will be shown later in Sec. \ref{sec:IT-E}. 
If the group eigenspaces are linearly independent each other, we can exactly separate (orthogonalize) the group eigenspaces by using the block diagonalization pre-beamforming. However, we assume mutual orthogonality between group eigenspaces, which is a stronger condition than linear independence, for the sake of ease of analysis in this work. We also let $T=1$ unless otherwise is mentioned, since each class consumes orthogonal resources. It is further assumed that $G$ groups are formed in a symmetric manner such that each group has the same number $K'=K/G$ of users and the same rank $r=M/G$ of $\pR_g = \pU_{g}\Lambdam_g\pU_g^\ct$ for simplicity, where $G$ divides both $K$ and $M$. It is not difficult to extend to the general case of multiple classes and asymmetric per-group parameters with orthogonality replaced by linear independence. 
In the sequel, we present an ideal structure of the transmit correlations taken from \cite{Nam12}:  

\begin{defi}[Unitary Structure] \label{def-1}
We say that a set of user groups has unitary structure of their transmit correlations  if all users in group $g$ have a common correlation eigenspace spanned by the tall unitary matrix 
$\pU_{g}$, and if the $M \times rG$ matrix $\underline{\pU} = [\pU_1,\cdots,\pU_G]$ is {unitary} such that $ \underline{\pU}^\mathsf{H} \underline{\pU}=\underline{\pU} \underline{\pU}^\ct = \Id.$
\end{defi}

For the general case of $rG\le M$, this structure is called tall unitary such that $ \underline{\pU}^\mathsf{H} \underline{\pU}= \Id$.
Under the unitary structure, we just let $b_g = r$ and $\Bm = \underline{\pU}$. These choices eliminate interference between $G$ groups and the 
resulting MIMO BC is given by
\begin{align} \label{eq:SM-5}
   \py_g & = \textsf{\pH}_g^\mathsf{H} \pP_g\pd_g +\pz_g = \Wm_g^\mathsf{H} \Lambdam_g^{1/2} \pP_g\pd_g +\pz_g
\end{align}
where $\Wm_g$ is an $r \times K'$ matrix with i.i.d. elements $\sim \mathcal{CN}(0,1)$, for all $g$, thereby yielding the reduced column dimensionality 
of the effective channel $\textsf{\pH}_g$. Using (\ref{eq:SM-5}), we arrive at the following simple yet important result \cite[Thm. 1]{Nam12}: 
Under the unitary structure, the ergodic sum capacity of the original MIMO BC
(\ref{eq:SM-3}) with perfect CSI is equal to that of parallel subchannels (\ref{eq:SM-5}) with reduced dimensional $\textsf{\pH}_g$, given by
\begin{align} \label{eq:MU-0}
 \sum_{g=1}^G \mathbb{E} \bigg[\max_{\sum_{g=1}^G\trace(\pS_g)\le P} \; \log  \Big| \Id + \Lambdam_g^{1/2} \Wm_g \Sm_g \Wm_g^\mathsf{H} \Lambdam_g^{1/2}  \Big|\bigg]
\end{align}
where $\Sm_g$ denotes the diagonal $K' \times K'$ input covariance matrix for group $g$ in the dual multiple-access channel (MAC).
This can be intuitively verified by noticing that the effective channel $\textsf{\pH}_g$ with reduced dimension of $K'\times r$ is unitarily equivalent to the original channel $\pH_g$ of $K'\times M$ under the unitary condition so that we can get \emph{effective channel dimension reduction} without loss of optimality. 
This dimension reduction effect provides significant savings in CSI uplink training (for TDD systems) and both in downlink training and CSI feedback (for FDD systems) 
by a factor of $G$. 

\section{Impact of Transmit Correlation to the Power Gain of MIMO BC}
\label{sec:IT}

It was shown in \cite{Nam13a} that, in the large number of users regime, transmit correlation may significantly help the capacity of Gaussian MIMO BCs. One may think that if we can fully exploit the ideal unitary structure, we might do better than the independent fading case also in other regimes of interest. Assuming perfect CSI throughout this section, we will show that this holds true in some cases but not in all cases of interest. This is due to the fact that there is a tradeoff between power loss resulting from the effective channel dimension reduction and beamforming gain from pre-beamforming across the long-term eigenspaces of the user groups. In order to understand the tradeoff, we carefully investigate the impact of transmit correlation on the power gain of the Gaussian MIMO BC for different regimes in terms of $r=M/G$, $K'=K/G$, and $G$. 

In the sequel, we first consider the asymptotic capacity bounds of correlated fading MIMO-BCs in the high-SNR regime and then characterize the high-SNR capacity in the large $M$ regime in a compact form. We also compare these results with the corresponding independent fading case in order to see if there exist potential benefits of transmit correlation to the power gain of the channels.

\subsection{High-SNR Analysis}
\label{sec:IT-A}


For $M$ fixed, we investigate the ergodic capacity bounds of MIMO BC at high SNR to capture the power offset between the i.i.d. Rayleigh fading channel and the correlated Rayleigh fading channel satisfying the unitary structure. 
Since a closed-form characterization of the ergodic sum capacity of MIMO BCs is very little known even in the perfect CSI (i.e., perfect CSI at the transmitter (CSIT) as well as at the receivers) case \cite{Gol03}, we rely on some upper and lower bounds.
Using the result in (\ref{eq:MU-0}), the well-known high-SNR equivalence between MIMO point-to-point channel and MIMO BC \cite{Cai03,vishwanath2003duality} (referred to as asymptotic point-to-point equivalence in this paper), and random matrix theory in \cite[Thm. 2.11]{Tul04}, we get the following bounds on the high-SNR capacity behaviors of correlated fading MIMO BCs.

\begin{thm} \label{thm-2}
Suppose perfect CSIT on ${\textsf{\pH}}_g$ in (\ref{eq:SM-10}) and the unitary structure of the users channel covariances as in Definition \ref{def-1}. 
For $r<K'$, the high-SNR capacity of the corresponding MIMO BC with correlated Rayleigh fading satisfies
\begin{align}  \label{eq:IT-13}
  M\log \frac{r}{K'}+o(1) \; \le \; \mathcal{C}^\text{sum}(P) -  \Big(M \log\frac{P}{M}+\sum_{g=1}^G\log|\Lambdam_g|+\kappa(K',r)\Big) \; \le \; o(1)
\end{align}
where $\kappa(x,y)= yG \big(-\gamma +\sum_{\ell=2}^{x}\frac{1}{\ell}+\frac{x-y}{y}\sum_{\ell=x-y+1}^{x}\frac{1}{\ell} \big)\log e$ with $\gamma$ the Euler-Mascheroni constant, $o(1)$ goes to zero as $P\rightarrow \infty$.
For $r\ge K'$, we have
\begin{align}  \label{eq:IT-14}
  \sum_{g=1}^G\sum_{i=1}^{K'}\log\frac{\lambda_{g,r-i+1}}{G} +o(1) \; \le \; \mathcal{C}^\text{sum}(P)  -  \Big(K \log\frac{P}{K'}+\kappa(r,K')\Big) \; \le \; o(1)
\end{align} 
where $\lambda_{g,i}$ is the $i$th diagonal element of $\Lambdam_g$.
\end{thm}

\begin{IEEEproof} 
See Appendix \ref{app:proof-2}.
\end{IEEEproof}

The above result can be generalized to the tall unitary structure for which $M \ge rG$ and $M$ in (\ref{eq:IT-13}) is replaced by $rG.$  
The lower bound in (\ref{eq:IT-13}) for the $r< K'$ case may become rather loose when $r\ll K'$. This is because the inherent multiuser diversity in the case of a very large number of users cannot be captured by the asymptotic point-to-point equivalence used to prove the lower bound (see (\ref{eq:IT-35b}) in Appendix  \ref{app:proof-2}).
Nevertheless, we will use these bounds in the sequel since they remain fairly tight as long as $K'/r$ is small. The upper bound in (\ref{eq:IT-13}) becomes asymptotically tight when the receivers \emph{inside} each group\footnote{Since users in a particular group are often closely located, the intra-group cooperation within such a group is more feasible than the full cooperation across all users over the entire BS coverage.} are allowed to cooperate, which we call \emph{intra-group cooperation} in this work. 
In the case of $r=K'$ (i.e., $M=K$), (\ref{eq:IT-13}) coincides with (\ref{eq:IT-14}) and it is asymptotically tight. In addition, if we relax the unitary structure as the tall unitary structure where $M \ge rG$,  (\ref{eq:IT-13}) becomes tight at high SNR for $r=K'$ but $M\ge K$ as well. 

\begin{rem}
An alternative expression of the asymptotic capacity behavior for $r\ge K'$ can be found by utilizing the approach in \cite{Shi03,Loz05}\footnote{Although these point-to-point results assume that only the distribution of a channel  is accessible at the transmitter, the difference from the perfect CSIT case that we are assuming vanishes at high SNR when the number of receive antennas is greater than or equal to the total number of transmit antennas (this is the case of the dual MAC in (\ref{eq:MU-0}) when $r\ge K'$).}.
Comparing with the alternative characterization and other previous results \cite{Ver05,Tul05} for the point-to-point MIMO case, we can see that (\ref{eq:IT-14}) in Theorem \ref{thm-2} is more intuitive and insightful. For example, (\ref{eq:IT-14}) will be used in Sec. \ref{sec:IT-D} to show that, for $r \ge K'$, in general we cannot do better than the independent fading case. Moreover, our result reveals that the impact of transmit correlation diversity to the high-SNR capacity in fact depends only on the non-zero eigenvalues of $\pR_g$. This is also the case in the point-to-point case where $G=1$.
\end{rem}

It immediately follows from \cite{Loz05} and \cite{Cai03} that, for $M\ge K$, the high-SNR capacity of the i.i.d. Rayleigh fading MIMO BC with perfect CSI behaves like
\begin{align}  \label{eq:IT-50}
  \mathcal{C}^\text{sum}_\text{iid}(P) = K \log\frac{P}{K}+\kappa(M,K)+o(1).
\end{align}

In what follows, we focus on the $r\ge K'$ case to investigate the impact of transmit correlation diversity   compared to the independent case. To do so, we will make use of the affine approximation introduced by Shamai and Verd\'u \cite{Sha01}. The high-SNR capacity 􏱖􏰹􏰳$\Cc(P)$ 􏰺 is well approximated by 􏰺the zero-order term in the expansion of the capacity as an affine function of SNR ($P$)
\begin{align}  \label{eq:IT-51}
  \Cc(P) = \Sc^\infty (\log P - \Lc^\infty) +  o(1)
\end{align}
where 􏱔􏰾$\Sc^\infty= \lim_{P\rightarrow \infty} \frac{\Cc(P)}{\log P}$ is the multiplexing gain and $\Lc^\infty= \lim_{P\rightarrow \infty} \big(\log P - \frac{\Cc(P)}{\Sc^\infty}\big)$ 􏱦􏰾is the power offset.
Using the quantity $\Lc^\infty$, for $r\ge K'$, we consider the difference between  $\Lc^\infty_\text{iid}$  and $\Lc^\infty_\text{ub}$, where $\Lc^\infty_\text{iid}$ is the power offset of (\ref{eq:IT-50}) and $\Lc^\infty_\text{ub}$ is the power offset of the upper bound in (\ref{eq:IT-14}) denoted by $\Cc^\text{sum}_\text{ub}(P)$, as shown by 
\begin{align}  \label{eq:IT-52}
   \Lc^\infty_\text{iid} - \Lc^\infty_\text{ub} & = \lim_{P\rightarrow \infty} \bigg(\frac{\Cc^\text{sum}_\text{ub}(P)}{\Sc^\infty} -\frac{\Cc^\text{sum}_\text{iid}(P)}{\Sc^\infty}\bigg) \nonumber \\
   &= \underbrace{\frac{3}{K}\sum_{g=1}^G \sum_{i=1}^{K'}\log \lambda_{g,i}}_\text{eigen-beamforming gain} + \underbrace{\frac{3}{K}\big(\kappa(r,K')-\kappa(M,K)\big)}_\text{channel dimension loss}.
\end{align}
where $3\approx 10\log_{10}2$ is due to the fact that the power offset $\Lc^\infty$ is in units of dB (i.e., horizontal offset in capacity versus SNR curves).
While the first term in (\ref{eq:IT-52}) can be a positive constant with $P$ depending on the degrees of transmit correlation diversity and the condition number of $\Lambdam_g$, the second term is only non-positive. Since the former (along with $\frac{3}{M}\sum_{g=1}^G\log|\Lambdam_g|$ in (\ref{eq:IT-13}) for $r<K'$) indicates the power gain due to pre-beamforming along group eigenspaces inherent in the unitary structure, we call this \emph{eigen-beamforming gain}. The latter will be referred to as \emph{channel dimension loss} as the channel dimension reduction in (\ref{eq:SM-5}) incurs such a loss in power offset.
As a result, transmit correlation diversity turns out to yield power loss as well as power gain. We also observe a \emph{tradeoff} between the eigen-beamforming gain and the channel dimension loss as $G$ is inversely proportional to $r$ for $M$ fixed.


Let us first consider the case of $r=K'$ (i.e., $M=rG=K$), in which the high-SNR behavior in (\ref{eq:IT-14}) reduce to 
\begin{align}  \label{eq:IT-14b}
  \mathcal{C}^\text{sum}(P)   &=  M \log\frac{P}{M}+M\Bigg(-\gamma +\sum_{\ell=2}^{r}\frac{1}{\ell} \Bigg)\log e+\sum_{g=1}^G\log|\Lambdam_{g}|+o(1).
\end{align} 
In order to investigate the tradeoff and easily evaluate the difference in (\ref{eq:IT-52}), we can upper-bound the asymptotic capacity behavior in (\ref{eq:IT-14b}) by letting 
\begin{align}  \label{eq:ID-25a}
  \lambda_{g,i}=\frac{M}{r}=G
\end{align} 
for all $(g,i)$ (For details on this upper bound, see (\ref{eq:IT-28b}) in Appendix \ref{app:proof-2} and (\ref{eq:ID-20}) in Appendix \ref{app:proof-8}). Then, the eigen-beamforming gain in (\ref{eq:IT-52}) is upper-bounded by 
$3\log G$.  
Using the approximation of the Harmonic number \cite{Con96}
\begin{align} 
   \sum_{\ell=1}^{n}\frac{1}{\ell} &=\gamma +\ln n +\sum_{m=2}^\infty \frac{\zeta(m,n+1)}{m} \nonumber \\
   &= \gamma +\ln n +\frac{1}{2n} -\frac{1}{12n^2} +\frac{1}{120n^4} +O({n^{-6}}) \nonumber
\end{align} 
where $\zeta(\cdot)$ is the Hurwitz zeta function, we have 
\begin{align}  \label{eq:IT-20}
   \Lc^\infty_\text{iid} - \Lc^\infty_\text{ub}& \le 3\bigg(\frac{1}{2r} -\frac{1}{12r^2} -\frac{1}{2M} +\frac{1}{12M^2} \bigg)\log e +O(M^{-4})\nonumber \\
   & = 3\bigg(\frac{G-1}{2M} -\frac{G^2-1}{12M^2}  \bigg)\log e +O(M^{-4}) \nonumber \\
   &\le \frac{3}{2r}\log e \; \text{ dB.}
\end{align}
This shows that, assuming the optimistic condition (\ref{eq:ID-25a}) on the condition number of $\Lambdam_g$, the power gain can be positive but marginal.  By investigating the $r>K'$ case in a similar way with some manipulations, we can see that $\Lc^\infty_\text{iid} - \Lc^\infty_\text{ub} \le \frac{3}{M}\big(\frac{G-1}{2} -\frac{G^2-1}{12M} -\sum_{\ell=r-K'+1}^r\frac{1}{\ell^2}\big)\log e+O(M^{-4})$, implying that even the marginal gain diminishes for $r> K'$. Accordingly, transmit correlation can be detrimental to the capacity depending on the condition number of $\Lambdam_g$.
The eigen-beamforming gain is shown to be almost completely offset by the power  loss due to the channel dimension reduction in this case. As a consequence, transmit correlation diversity in general provides no capacity gain for the unitary structure with $r\ge K'$. 

\subsection{Large $K$ Analysis}
\label{sec:IT-D}

We first consider the case where $r$ is not significantly larger than $K'$. 
If the intra-group cooperation is allowed, the high-SNR capacity of correlated fading BCs can approach the corresponding i.i.d Rayleigh fading point-to-point case, depending again on the condition number of $\Lambdam_{g}$. In contrast, the independent fading BC case needs the full cooperation to achieve the same high-SNR capacity. But, this seems infeasible and the corresponding channel is not a BC any more. As a result, transmit correlation diversity is beneficial at least in this sense for $r< K'$ but not $r \ll K'$.

The sum-rate scaling in the large $K$ regime where $r \ll K'$, relevant in practice for hot-spot scenarios, was already addressed in \cite{Nam13a}, but without sufficient exposition. 
It is well known from Sharif and Hassibi \cite{Sha05} that the sum capacity of the i.i.d. Rayleigh fading MIMO Gaussian BC scales like
$$  \Cc^\text{sum}_{\rm iid}(K) = M\log\frac{P}{M}+M\log\log K +o(1).$$
In the special case where all users have both the same SNR and the common transmit correlation matrix $\pR$ of full rank, the authors in \cite{Aln09} proved that the sum capacity scales like
$  M\log\frac{P}{M}+M\log\log K +\log|\pR|+o(1)$
where $\log|\pR|\le 0$ due to $\trace(\pR)=M$. 

Notice that the case where all users have the same correlation corresponds to the case of $G = 1$, i.e., the poorest case of transmit correlation diversity. 
As a matter of fact, for $G > 1$ under the unitary structure where different groups of users have orthogonal eigenspaces, it was shown in \cite{Nam13a} that,  
for fixed $M$ and large $K$, the asymptotic sum capacity of correlated  Rayleigh fading MIMO BC is
\begin{align}  \label{eq:TC-3}
  \mathcal{C}^\text{sum}(K)=M\log\frac{P}{M}+M\log\log K +\sum_{g=1}^G\log|\Lambdam_{g}|+o(1)
\end{align}
where the detailed achievability proof is given in Appendix \ref{app:proof-3}. This shows that, for the large $K$ regime with correlated fading, there exists an additional term thanks to the eigen-beamforming gain as well as the well-known multiuser diversity gain term $M\log\log K$. 
As an upper bound on the potential gain of transmit correlation diversity in the $r\ll K'$ regime, 
if the AS of group $g$, $\Delta_g$, is close to zero but Rayleigh fading is still valid,
then $$ \limsup_{\Delta_g\rightarrow 0, \forall g}\Lc^\infty_\text{iid}(K) - \Lc^\infty(K) = 3\log M$$ 
where $\Lc^\infty_\text{iid}(K)$ and $\Lc^\infty(K)$ are the power offsets of $\Cc^\text{sum}_{\rm iid}(K)$ and $\Cc^\text{sum}(K)$, respectively. 

In the case of $r\ge K'$, the eigen-beamforming gain could be completely compensated by the channel dimension loss, yielding that transmit correlation does not help the capacity. In the $r\ll K'$ case, however, the channel dimension loss\footnote{In this case, the channel dimension loss can be interpreted as multiuser diversity  reduction with respect to the independent fading case due to the fact that user selection is independently performed in a group basis for only $K'$ users rather than $K$ users.} \emph{vanishes} in the large $K$ regime as shown in Appendix \ref{app:proof-3}, while eigen-beamforming can still provide the power gain of up to $3\log G$ dB, which can be translated into the rate offset of $M\log G$ bps/Hz. This explains why the correlated fading case can significantly outperform the independent fading case in this regime. 

Finally, we consider the intra-group cooperation for large $K'$ but not necessarily $r\ll K'$ and compare its performance with (\ref{eq:TC-3}).
\begin{corol} \label{cor-3}
Assuming the intra-group cooperation between the receivers within each group, we have
\begin{align}  \label{eq:IT-41}
  \Cc^\text{sum}(P,K) = M\log\frac{P}{M}+M\log K' +\sum_{g=1}^G\log|\Lambdam_{g}|+o(1).
\end{align}
\end{corol}

This can be verified with the high-SNR upper bound in (\ref{eq:IT-37a}) in Appendix \ref{app:proof-2} and the fact that $   \mathbb{E} \big [\log\left|  \Wm_g \Wm_g^\ct \right|\big ]  \simeq r \log K'$ for large $K'$, which follows from (\ref{eq:ID-10}). The intra-group cooperation is shown to provide the additional power gain of $3\big(\log K'-\log \log K\big)$ dB at high SNR for large $K'$, compared to (\ref{eq:TC-3}) without cooperation.

\subsection{Large System Analysis}
\label{sec:IT-B}

We turn our attention to the large number of antennas regime, i.e., the large system analysis. For this analysis, we need the asymptotic behavior of large-dimensional Wishart matrices. To this end, a common approach utilizes known results from asymptotic random matrix theory \cite{Tul04} (e.g., see \cite{Ver99} based on the Mar\v{c}enko-Pastur law \cite{Mar67}). In this paper, we shall instead consider direct analysis of the asymptotics of the capacity bounds in Theorem 1.

Let $$\mu=\frac{M}{K}=\frac{r}{K'}$$ and $G$ be fixed such that both $r$ and $K'$ are taken to infinity along with $M$.

\begin{thm} \label{thm-8}
Suppose the perfect CSIT on ${\textsf{\pH}}_g$, the unitary structure, and the uniform boundedness of $\lambda_{g,i}$ such that 
\begin{align}  \label{eq:SM-1c}
  0 < \epsilon \le \frac{\lambda_\text{min}}{\lambda_\text{max}} \le 1
\end{align}   
for all $g$ and any $i \in \mathbb{Z^+}$.  As $M\rightarrow \infty$, for $\mu < 1$,  the high-SNR capacity of the corresponding correlated fading MIMO BCs scales linearly in $M$ with the ratio
\begin{align}  \label{eq:ID-5}
  \log \frac{\mu\lambda_\text{min}}{G} +o(1)\; \le \; \lim_{M\rightarrow \infty} 
  \frac{\mathcal{C}^\text{sum}(P,M,r)}{M} - \log \frac{P}{e\mu}+\Big(\frac{1-\mu}{\mu}\Big)\log\frac{1}{1-\mu} \; \le \; o(1)
\end{align}
where $o(1)$ is a constant with $M$ but vanishes as $P\rightarrow \infty$.

For $\mu \ge 1$, the high-SNR capacity scales linearly in $K$ with the ratio
\begin{align}  \label{eq:ID-6} 
   \log \frac{\lambda_\text{min}}{G} +o(1)\; \le \; \lim_{M\rightarrow \infty} \frac{\mathcal{C}^\text{sum}(P,M,r)}{K}  &- \log \frac{\mu P}{e}+(\mu-1)\log\frac{\mu}{\mu-1} \; \le \; o(1).
\end{align}
\end{thm}

\begin{IEEEproof} 
See Appendix \ref{app:proof-8}.
\end{IEEEproof}

When $M=K$ with $\pR_g=\pI_r$ for all $g$, we can see that  (\ref{eq:ID-6}) reduces to 
\begin{align} \label{eq:ID-23}
   \frac{\Cc^\text{sum}(P,M)}{M} =\log \frac{P}{e}+o(1)
\end{align}
which equals the well-known capacity scaling of the i.i.d. Rayleigh fading MIMO channel \cite{Fos96}. It immediately follows from Theorem \ref{thm-8} that the asymptotic capacity scaling of the i.i.d. Rayleigh fading MIMO BC is upper-bounded by 
\begin{align}  
  \frac{\mathcal{C}^\text{sum}_\text{iid}(P,M,r)}{M} &= \log \frac{P}{e\mu}+\Big(\frac{1-\mu}{\mu}\Big)\log\frac{1}{1-\mu} +o(1), \ \text{if } \mu< 1 \label{eq:ID-5b} \\
  \frac{\mathcal{C}^\text{sum}_\text{iid}(P,M,r)}{K}  &= \log \frac{\mu P}{e}+(\mu-1)\log\frac{\mu}{\mu-1} +o(1), \hspace{2.5mm} \text{if } \mu\ge 1 \label{eq:ID-6b}.
\end{align}
These are also the upper bounds of the point-to-point case. In particular, (\ref{eq:ID-5b}) for $\mu< 1$ is the same as \cite[Proposition 2]{Loz05}.
We can see that the growth rates of the capacity of correlated fading channels under the unitary structure are upper-bounded by the i.i.d. Rayleigh fading case in (\ref{eq:ID-5b}) and  (\ref{eq:ID-6b}) irrespectively of $\mu$. Although the additional power gain of up to $3\log G$ dB in (\ref{eq:TC-3}) in the large $K$ regime may seem to contradict the observation from Theorem \ref{thm-8}, the assumption therein is different in that only $K$ increases to infinity with $M$ fixed, while both $M$ and $K$ here increase with a fixed ratio $\mu<1$. 

The assumption of the uniform boundedness of non-zero eigenvalues $\lambda_{g,i}$ of $\pR_g$ may seem unrealistic since $\pR$ is generally of full algebraic rank even if eigenvalues except dominant ones decay quickly. However, it is quite reasonable at least in the large number of antennas regime with the antenna configuration of ULA. For this case, it was shown in \cite{Adh13} that non-zero eigenvalues of $\pR$ can be accurately approximated by a set of samples $\{S([m/M]): m=0,\cdots,M-1\}$ (with $[x]$ being $x$ modulo the interval $[-1/2,1/2]$) which has support of length $\rho\le 1$ on such an interval. Here $S(\cdot)$ is the eigenvalue spectrum (discrete-time Fourier transform) of $\pR$. 
This implies that non-dominant eigenvalues go to zero when $M$ is sufficiently large. In realistic channels, $r$ should be considered as an \emph{effective rank} denoting the number of dominant eigenvalues \cite{Nam12}.

For the sake of concreteness, in this paper, we will consider the one-ring model for $\pR$, which corresponds to the typical cellular downlink case where the BS is elevated and free of local scatterers, and the user terminals are placed at ground level and are surrounded by local scatterers. 
In the one-ring model, a user located at azimuth angle $\theta$ and distance ${\sf s}$ is surrounded by a ring of scatterers of radius  ${\sf r}$ such that angular spread (AS) $\Delta \approx \arctan({\sf r}/{\sf s})$. Assuming the ULA with a uniform distribution of the received power from planar waves impinging on the BS array, the correlation coefficient between BS antennas $1 \leq p, q \leq M$ is given by
\begin{align} \label{eq:SM-4}
   [\pR]_{p,q} =\frac{1}{2\Delta}  \int_{-\Delta}^{\Delta}e^{j2\pi D(p-q)\sin(\omega+\theta)}d\omega
\end{align}
where $D$ is the normalized distance between antenna elements by the wavelength. 


In order to better understand the asymptotic behaviors of $\frac{1}{M}\sum_{g=1}^G\log \left |\Lambdam_g \right |$ and hence to obtain tighter bounds on the capacity scaling in (\ref{eq:ID-5}) and (\ref{eq:ID-6}), we may utilize the following approximation.
Assuming the ULA antenna, the transmit correlation matrix $\pR_g$ of group $g$ in (\ref{eq:SM-4}) can be given in a Hermitian Toeplitz form.
The eigenvalue spectrum $S(\xi)$ of $\pR_g$ is defined by the discrete-time Fourier transform of the coefficients $r_k\triangleq[\pR_g]_{\ell,\ell-k}$, i.e., 
$$S(\xi)= \sum_{k=0}^\infty r_k e^{-j2\pi k\xi}, \ \xi \in [-1/2,1/2].$$
For most cases of interest, this is a uniformly bounded absolutely integrable function over $\xi$. Then, the limiting behavior of $\frac{1}{r}\log|\pR_g| $ can be explicitly expressed by using the well-known Szeg\"{o} theorem \cite{Gre84, Gra06} 
as follows: 
\begin{align} \label{eq:ID-29}
   \lim_{r\rightarrow \infty}\frac{1}{r}\log|\pR_g| = \int_{-1/2}^{1/2} \log S(\xi) d\xi .
\end{align}
For example, for the one-ring scattering model considered in \cite{Adh13} that the eigenvalue spectrum $S(\xi)$ can be characterized in terms of only the geometric channel parameters as 
\begin{align} \label{eq:ID-30}
    S(\xi) = \frac{1}{2\Delta} \sum_{k\in [D\sin(-\Delta+\theta)+\xi,\; D\sin(\Delta+\theta)+\xi]} \frac{1}{\sqrt{D^2-(k-\xi)^2}}. 
\end{align}
In general, we can accurately predict $\frac{1}{r}\log|\Lambdam_g|$ thanks to (\ref{eq:ID-29}) from the scattering geometry that characterizes the propagation between a user group and the base station antenna array, avoiding the need for the eigendecompsition of the large-dimensional matrix $\pR_g$.

\subsection{Numerical Results and Summary}
\label{sec:IT-E}

\begin{figure}
\vspace{-3mm}
\center  \includegraphics[scale=.7]{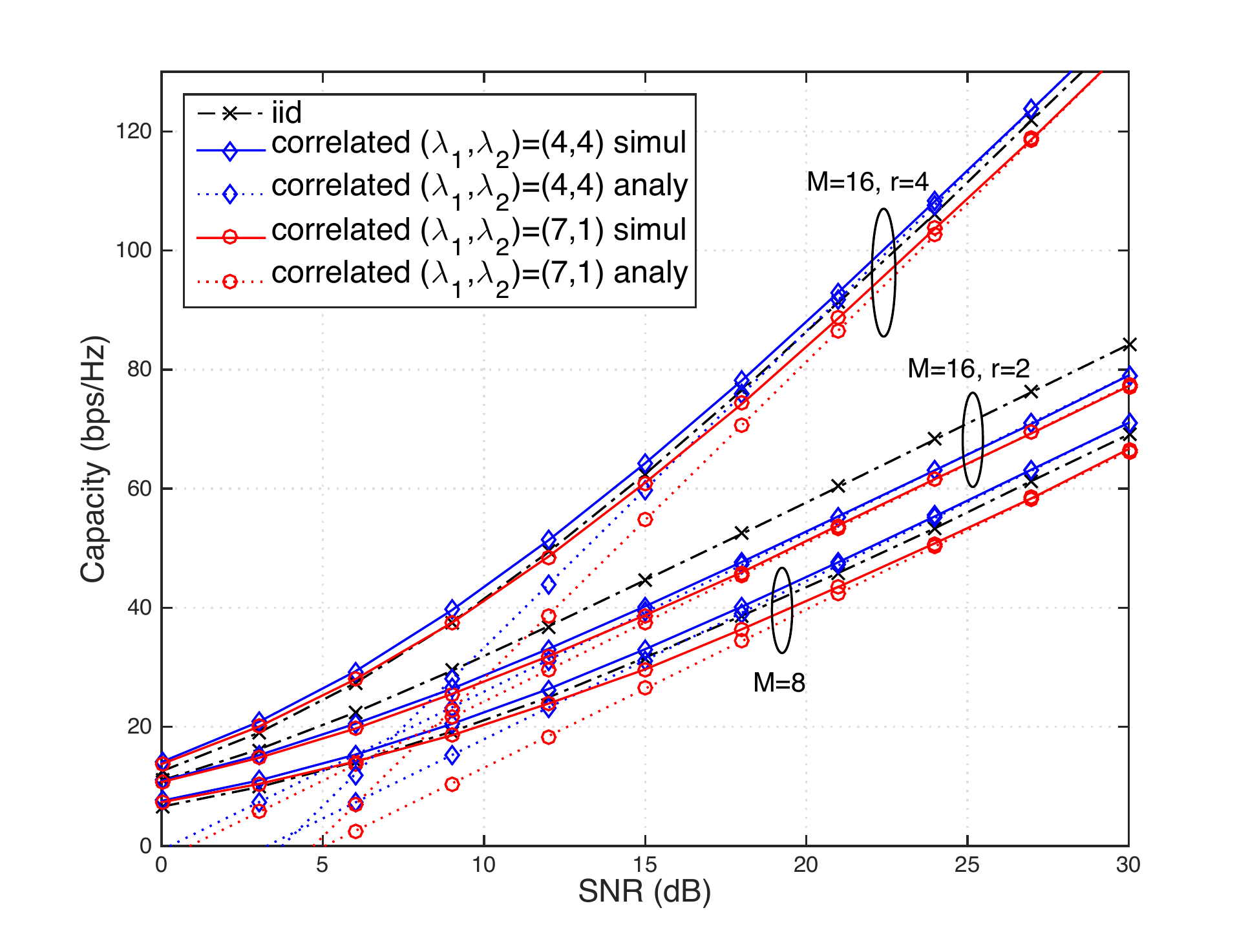}
  \caption{Sum capacity vs. SNR curves for $M=8,16$ with all satisfying $G$ degrees of transmit correlation diversity. ``analy" and ``simul" represent the high-SNR sum-rate scaling in Theorem \ref{thm-2} and Monte Carlo simulation, respectively. For $M=K=8$ and $r=2$, $(\lambda_1,\lambda_2)$ is given in the legend. For $M=16, K=8$ and $r=2$, the blue curves indicate $(\lambda_1,\lambda_2)=(8,8)$, while the red curves $(\lambda_1,\lambda_2)=(12,4)$. Finally for $M=K=16$ and $r=4$, the blue curves indicate $(\lambda_1,\lambda_2,\lambda_3,\lambda_4)=(4,4,4,4)$, while the red curves $(\lambda_1,\lambda_2,\lambda_3,\lambda_4)=(7,5,3,1)$.}\label{fig-3a}
\vspace{-2mm}
\end{figure}

This subsection provides some numerical results to show the validity of the impact of transmit correlation diversity to the capacity in the previous subsections.  In order to generate a unitary structure, a set of the group eigenvector matrices $\{\pU_g\}_{g=1}^G$ were obtained by a randomly chosen channel covariance matrix, since the arbitrary realizations of the eigenvector matrices do not change the capacity as long as $\pU_g^\ct\pU_{g'}=\pI$ for all $g'\neq g$. For this structure, we assume $\Lambdam_g=\Lambdam$ for all $g$ for convenience, where $\Lambdam=\diag(\lambda_1,\cdots,\lambda_r)$.

Fig. \ref{fig-3a} depicts the ergodic sum capacity versus SNR curves of the i.i.d. and the correlated Rayleigh fading MIMO BCs for different $M, K,$ and $\Lambdam$ with $G=4$. In the unitary structure with $M=K$ (i.e., $r=K'$), whether transmit correlation diversity can be beneficial to the capacity or not depends on the condition number of $\Lambdam$, as mentioned earlier. Moreover, the high-SNR sum-rate bounds in Theorem \ref{thm-2} are shown to be tight when $M=K$.  For $M=16, K=8$ and $r=2$, we can see that our asymptotic bound is also tight for the tall unitary structure with $M> K$ and $r=K'$ and that the tall unitary structure suffers from performance degradation relative to the unitary structure.

\begin{figure}
\vspace{-3mm}
\center  \includegraphics[scale=.7]{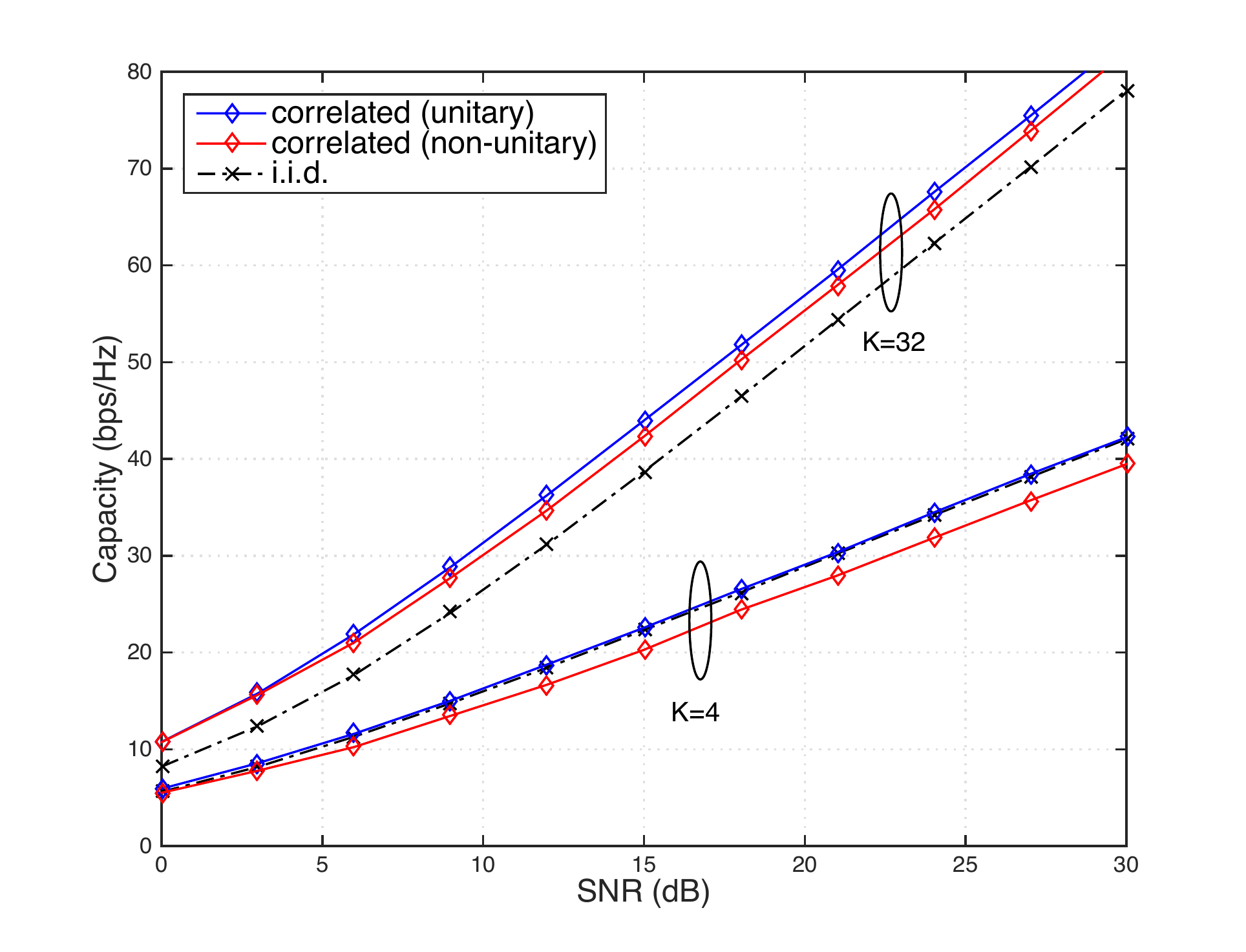}
  \caption{Sum capacity vs. SNR curves for $M=8$ with different numbers ($K$) of users. While $G=4$ for the unitary structure, $\theta_i \in [-60^o, 60^o]$ and $\Delta_i\in[5^o,10^o]$ for the non-unitary case. }\label{fig-4}
\vspace{-2mm}
\end{figure}

Fig. \ref{fig-4} also compares the capacities of the independent and correlated fading cases for different $K$ with $M=8$. For the unitary structure, we set $(\lambda_1,\lambda_2)=(4,4)$, i.e., the optimistic case in (\ref{eq:ID-25a}), for all $K$. The correlated fading case has a larger capacity than the independent fading case for $K=32$ ($M<K$), while  transmit correlation diversity does not help the capacity for $K=4$ ($M>K$). 
In addition, we would like to see if the foregoing results on the beneficial impact of transmit correlation derived by imposing the unitary structure are still valid for realistic channels not assuming the structure. To generate this ``non-unitary" (unstructured) case, we utilize the one-ring channel model in (\ref{eq:SM-4}) with the ULA of $D=1/2$ (half wavelength) where $\theta_i$ is uniformly distributed over the range of $[-60^o, 60^o]$ and $\Delta_i$ is uniformly distributed over the range of $[5^o,10^o]$, where $\theta_i$ and $\Delta_i$ are AoD and AS of user $i$, respectively. 
Then, we can see that, for $K=32$, transmit correlation is still noticeably beneficial to the capacity even if we did not assume the unitary structure. This remarkable result is also observed in the following evaluation. 

\begin{figure}
\vspace{-3mm}
\center  \includegraphics[scale=.7]{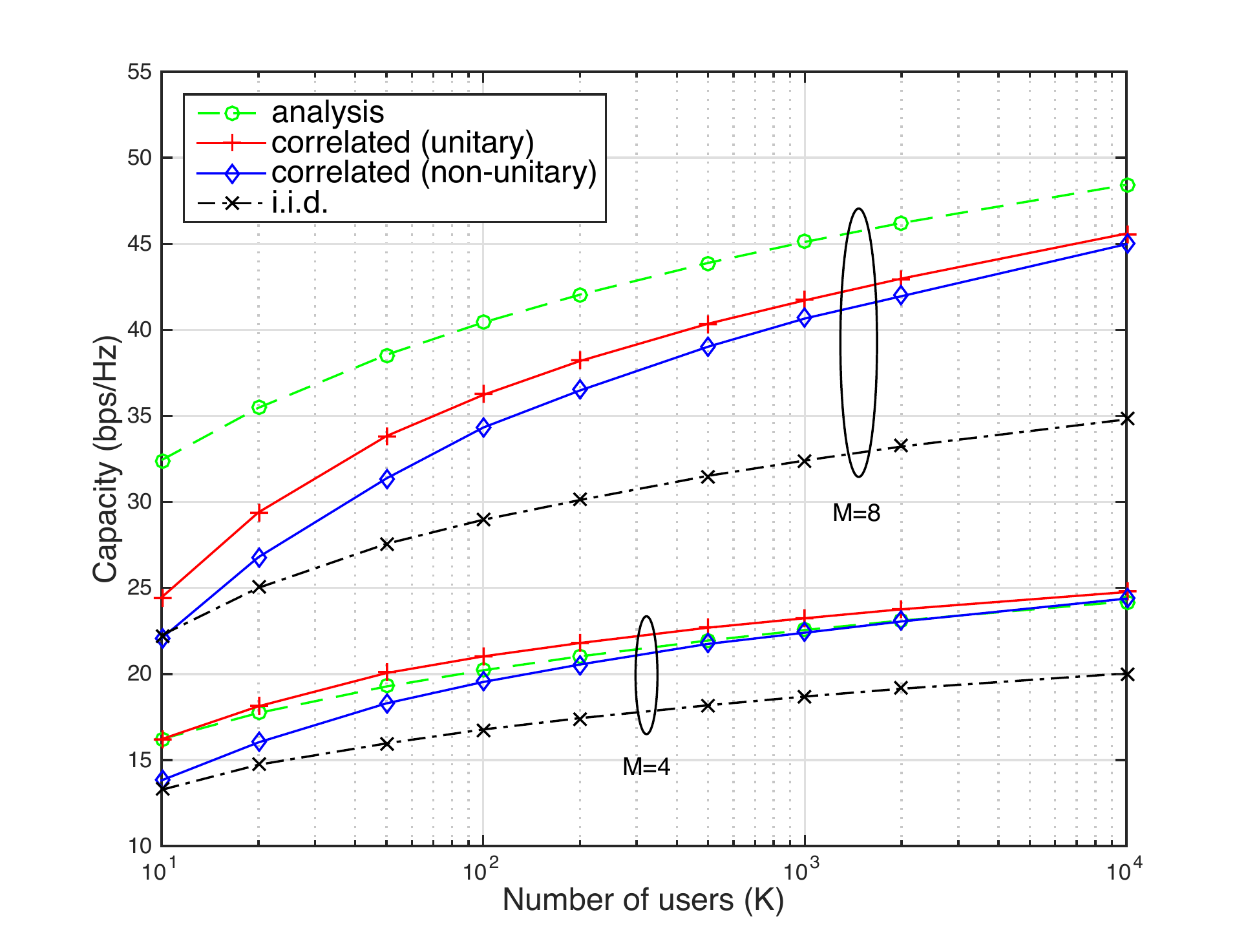}
  \caption{Sum capacity vs. the number of users for $M=4,8$ at SNR $= 10$ dB.}\label{fig-7}
\vspace{-2mm}
\end{figure}

Fig. \ref{fig-7} depicts the sum capacity versus the number of users curves for different $M$. For the unitary structure, we set $G=4$ for $M=8$ and $G=2$ for $M=4$ with $(\lambda_1,\lambda_2)=(4,4)$ for all $M$. The rate gap between correlated and independent fading cases gets larger as $K$ and $G$ increase. The exact capacity is shown to be predictable by the analysis in (\ref{eq:TC-3}) for $M=4$, while a much larger $K$ would be needed for the same asymptotic capacity in (\ref{eq:TC-3}) to converge for $M=8$. For the non-unitary case, we set $\Delta_i\in [5^o,20^o]$.
When $K=10,000$, the rate gap between the independent and non-unitary cases is  $4.3$ bps/Hz for $M=4$ where the potential gain of $M\log G$ is $4$ bps/Hz, while it is $10.2$ bps/Hz for $M=8$ where $M\log G=16$. Therefore, a surprisingly large portion of the potential rate gain (i.e., power gain) of transmit correlation diversity is shown to be achievable for sufficiently large $K$ with the realistic setup where no  structure of users transmit correlations was assumed.

Assuming the unitary structure, the \emph{full-CSI} capacity results in this section can be summarized as follows.
\begin{itemize}
\item For $r\ge K'$ (i.e., $M=rG\ge K$), we can obtain the power gain of \emph{at most} $\frac{3}{2r}\log e$ dB (i.e., $\frac{G}{2}\log e$ bps/Hz rate gain) at high SNR, given the optimistic condition where the eigenvalues of $\pR_g$ are approximated by $\frac{M}{r}$.
Depending on the condition number of $\pR_g$, transmit correlation diversity is hence even detrimental to the capacity of MIMO BCs in this regime.

\item For $r<K'$ but not $r\ll K'$, 
our numerical results indicate that transmit correlation diversity yields a considerable capacity gain even when $K$ is not so larger than $M$. In addition, the intra-group cooperation is sufficient to achieve the full-CSI capacity of the point-to-point case. 
 
\item For $r \ll K'$, transmit correlation diversity can increase power gain by up to $3\log G$ dB over the i.i.d Rayleigh fading BC at any SNR. 

\item Although transmit correlation diversity is well defined for the unitary structure, its effect is {not} restricted to the structured case. It is observed from numerical results that a ``semi-unitary" structure is implicitly created by multiuser diversity (i.e., user selection/scheduling) for $r< K'$, yielding that even unstructured transmit correlations of users do help the capacity.  
\end{itemize}

It turns out that transmit correlation diversity may be rather detrimental to the power offset of MIMO BCs for $M\ge K$ except for some optimistic conditions, in contrast to the $M< K$ case. This is mainly due to the fact that the former case  suffers from the power loss due to effective channel dimension reduction, while such a loss vanishes owing to multiuser diversity in the latter as $K$ gets much larger than $M$.

So far, we have assumed prefect CSIT with \emph{no cost}, for which in general we cannot do better with transmit correlation diversity for $M\ge K$. Notice that the typical scenario of large-scale MIMO belongs to this unfavorable case which includes $M\gg K$. Consequently, transmit correlation diversity could not improve the performance of large-scale MIMO systems. It will be shown in the following section that this argument is not true for realistic pilot-aided systems, where CSIT is provided at the cost of downlink training.


\section{Fundamental Limits of Pilot-Aided MU-MIMO Systems}
\label{sec:FL}

In this section, we investigate asymptotic capacity bounds of pilot-aided MU-MIMO systems, in which the resources for downlink training are taken into consideration. While FDD systems make use of downlink common pilot and CSI feedback, TDD systems employ uplink dedicated pilot to exploit the uplink-downlink channel reciprocity  for downlink training. Both FDD and TDD need dedicated pilots for users to estimate their downlink channels for coherent detection \cite{Cai10}. Dedicated pilots go through the downlink beamforming vectors and hence they can be sent all together on the same time-frequency resource without significant interference, provided that $M$ is sufficiently large, so that they may consume a negligible resource. Therefore, we focus on the cost of the common downlink pilots in FDD, where the CSI feedback issue was already addressed in \cite{Nam12}, or the uplink per-user pilots in TDD.

In the independent fading channel, the downlink FDD system  uses $M_\text{iid}^*$ downlink dimensions to allow users to estimate the 
$M_\text{iid}^*$-dimensional channel vectors, where $M_\text{iid}^* = \min\{M,K,\lfloor T_c/2\rfloor\}.$ 
Assuming that CSIT is acquired by the base station through a delay-free and error-free feedback channel,\footnote{This is clearly an over-optimistic assumption, but it is in general good enough to obtain a simple bound
on what it is actually possible to achieve through realistic feedback implementations.} 
the high-SNR capacity of MU-MIMO downlink systems is upper-bounded by 
\begin{align}  \label{eq:intro-1}
  M_\text{iid}^*(1-M_\text{iid}^*/T_c)\log \snr+O(1).
\end{align}
If we optimize the number of (active) base station antennas and 
users as a function of the channel coherence time $T_c$, it turns out that 
\[ M_\text{iid}^*\left(1-\frac{M_\text{iid}^*}{T_c}\right) \leq \frac{T_c}{4}, \]
where the upper bound is attained by activating only $T_c/2$ antennas to serve $T_c/2$ users. 
Therefore, the system multiplexing gain does not \emph{scale} with $\min(M,K)$ and the per-user throughput vanishes as $O(\frac{1}{K})$, when $K$ becomes large and $T_c$ is fixed. 
This upper bound is also valid for TDD systems with reciprocity. 
For example, the number of scheduled users (specifically, $s=M_\text{iid}^*$ for the independent fading case) among the entire $K$ users is limited by uplink pilot 
overhead (also affected by $T_c$), which is optimized by letting $s = T_c/2$ and devoting half of the coherence block to uplink training (see \cite{Mar10}). 
If instantaneous feedback within coherence time $T_c$ is not possible, the impact of the resulting channel prediction error on the 
system multiplexing gain can be found in \cite{Cai10}.

As already pointed out, the above factor $T_c/2$ significantly limits the system performance for both $M$ and $K$ large. 
However, noticing that this result holds true in the \emph{independent fading} channel, we are to characterize some  performance 
limits in \emph{correlated fading} channels under the unitary structure in the following.


\subsection{Training Overhead Reduction}
\label{PS-0}

\subsubsection{FDD (Multiple pre-beamformed Pilot)}

The common pilot is in general isotropically transmitted, since it has to be seen by all users. We first consider a simple training scheme for FDD systems, where the downlink common pilot signal $\pX_g^\text{dl}$ for group $g$ is given by the pre-beamforming matrix $\pB_g$ as follows:
$$\pX_g^\text{dl}=\rho_\text{tr}\pB_g$$
where $\rho_\text{tr}$ indicates the power gap between the training phase and the communication phase. Assuming the unitary structure, we can just let $\pB_g=\pU_g$ and then the received pilot signal matrix for group $g$ is given by
\begin{align}  \label{eq:PS-1}
  \pY^\text{dl}_g = \pH_g^\ct \pX^\text{dl} +\pZ_g^\text{dl} = \rho_\text{tr}\textsf{\pH}_g^\ct +\pZ_g^\text{dl} 
\end{align}
where $\pX^\text{dl} =\sum_{g=1}^{G} \pX_g^\text{dl}$. This indicates that $G$ \emph{pre-beamformed pilot} signals $\px^\text{dl}_{g,i}$, $\forall g$, where $\px^\text{dl}_{g,i}$ is the $i$th column of $\pX^\text{dl}_{g}$, can be multiplexed and transmitted through a single pilot symbol and hence the overall common pilot signal $\pX^\text{dl}$ consumes only $r$ symbols, reduced by a factor of $G$.

Based on the above noisy observation of the pilot signal, each user in group $g$ can estimate the effective channel $\textsf{\ph}= \pU_g^\ct \ph$, which is unitarily equivalent to ${\ph}$ in (\ref{eq:SM-2}) under the unitary structure, as shown in Sec. \ref{sec:TCD}. Therefore, the proposed common pilot does not incur any loss due to pre-beamforming, as if it were a conventional pilot signal isotropic to all users.
A generalization of the above scheme was already given in \cite{Adh13}, which chose $\pX_g^\text{dl}=\pB_g\pU^\text{dl}$ with $\pU^\text{dl}$ being a scaled unitary matrix of size $r\times r$,  thereby making the downlink common pilot signal for each of antennas spread over $r$ pilot symbols. However, the previous work did not consider only an optimization of the system degrees of freedom taking into account the cost for downlink training dimension but also TDD or uplink systems in the next subsection.

\subsubsection{TDD}

The same idea above can be naturally applied to the TDD case with receive beamformer $\pU_g^\ct$ for the uplink dedicated pilot. To be specific, 
the received pilot signal matrix for TDD systems can be given by
$$\pY^\text{ul} = \sum_{g=1}^G \pH_g  \rho_\text{tr}\pI_{K'} +\pZ^\text{ul}. $$
By receive beamforming, i.e., multiplying from the left by $\pU_g^\ct$ for group $g$, we have
\begin{align}  \label{eq:PS-2}
  \pY^\text{ul}_g = \pU_g^\ct\pY^\text{ul} = \rho_\text{tr}\textsf{\pH}_g +{\pZ}_g^\text{ul} 
\end{align}
where ${\pZ}_g^\text{ul} =\pU_g^\ct{\pZ}^\text{ul}$.  The uplink dedicated pilot signal for all $K$ users consumes only $K'$ symbols, reduced by a factor of $G$ again. As a result, we can obtain the pilot saving not only in FDD systems but also in TDD systems, where the unitary structure is uplink-downlink reciprocal. Notice that such a pilot saving is also valid for MIMO MAC, i.e., MU-MIMO uplink systems. In \cite{Yin13}, a similar idea to the unitary structure was differently used to eliminate the pilot contamination effect in the  multi-cell TDD network instead of reducing the overhead for uplink dedicated pilot in each single cell. 


\subsection{Pilot-Aided System I}
\label{PS-1}

Assuming the unitary structure in a symmetric fashion such that $M=rG$ with the perfect knowledge on channel second-order statistics available at the transmitter, we can have up to $G$ groups with $r$ (long-term) eigenmodes each. However, using too many eigenmodes per group may degrade the performance of pilot-aided systems due to the cost of downlink training. Inspired by the pilot-aided system in \cite{Zhe02}, this section is devoted to maximize the system multiplexing gain for the pilot scheme in Sec. \ref{PS-0} with $T_c$ finite. 
Since we have to use all the $M$ antennas to preserve the unitary structure, we cannot directly follow the same line in \cite{Zhe02}. Rather, since it does not make sense to use less than $G$ degrees of transmit correlation diversity available in the MU-MIMO system, we just need to investigate how many eigenmodes (instead of active antennas) per group should be used in the communication phase. Although we focus on the downlink system in this section, the same multiplexing gain can be achieved in the uplink system according to (\ref{eq:PS-2}).

Suppose that we use $q$ of $r$ eigenmodes per group in the communication phase for FDD systems\footnote{In the TDD case, it suffices to suppose that we schedule $q$ of $K'$ users, where $q\le K'$, and to optimize the degrees of freedom with respect to $q$ taking into account the uplink pilot overhead.}, where $q\le r$.  Then, the number of degrees of freedom for communication within each group is upper-bounded by  
\begin{align} \label{eq-01}
   \min\{q,K'\}\left(T_c-q\right) 
\end{align}
where we devote only $q$ channel uses for the training phase thanks to the pilot scheme in Section \ref{PS-0}. We call this \emph{pilot-aided system I} in this work. The optimal number of eigenmodes per group to maximize (\ref{eq-01}) is given by
\begin{align} \label{eq-0}
   q^*=\min\left\{r,K',\Big\lfloor \frac{T_c}{2}\Big\rfloor\right\}
\end{align}
yielding the pre-log factor 
$q^*G\big(1-\frac{q^*}{T_c}\big)$. This indicates that there is no need for using more than $q^*$ eigenmodes per group. 
Therefore, we obtain the following results:
\begin{itemize}
\item Assuming the unitary structure with $G$ degrees of transmit correlation diversity, the high-SNR capacity of pilot-aided MU-MIMO systems is upper-bounded by
\begin{align} \label{eq:intro-4}
  M^*\left(1-\frac{M^*}{T_cG}\right) \log\snr + O(1) 
\end{align}
where $M^* = q^*G= \min\big\{M,K,\lfloor \frac{T_cG}{2}\rfloor\big\}$. 
\item Then, we have the fundamental limit on the system multiplexing gain
$$\lim_{\min\{M,K\}\rightarrow \infty} M^*\left(1-\frac{M^*}{T_cG}\right)=\frac{T_cG}{4}$$
for $T_cG\in 2\mathbb{Z}^+$. 
\end{itemize}

It turns out that, for both $M$ and $K$ large, exploiting $G$ degrees of transmit correlation diversity  can increase the system multiplexing gain by a factor of $G$, compared to the independent fading case. 
It is evident that as long as the degrees of transmit correlation diversity is sufficiently large such that $G\ge {2\min\{M,K\}}/{T_c}$ (i.e., $M^*=\min\{M,K\}$), the optimal number $M^*$ of eigenmodes is not affected any longer by the coherence time interval $T_c$. As a consequence, the system multiplexing gain is not saturated but rather it can keep growing as $\min\{M,K\}$ increases. 
If $M^*=M$ (or $K$), the high-SNR capacity of the pilot-aided system is (\ref{eq:IT-13}) (or (\ref{eq:IT-14})).
For $M^*=\lfloor \frac{T_cG}{2}\rfloor$, the high-SNR capacity equals to (\ref{eq:IT-14}) with $K$ and $K'$ replaced by $\lfloor \frac{T_cG}{2}\rfloor$ and $\lfloor \frac{T_c}{2}\rfloor$, respectively. 

The following example compares the upper bound (\ref{eq:intro-1}) on the system multiplexing gain for the independent fading case and the new upper bound (\ref{eq:intro-4}) for the correlated fading case, when $T_c$ is taken from real-life cellular systems.

\begin{figure} 
\center \includegraphics[scale=.8]{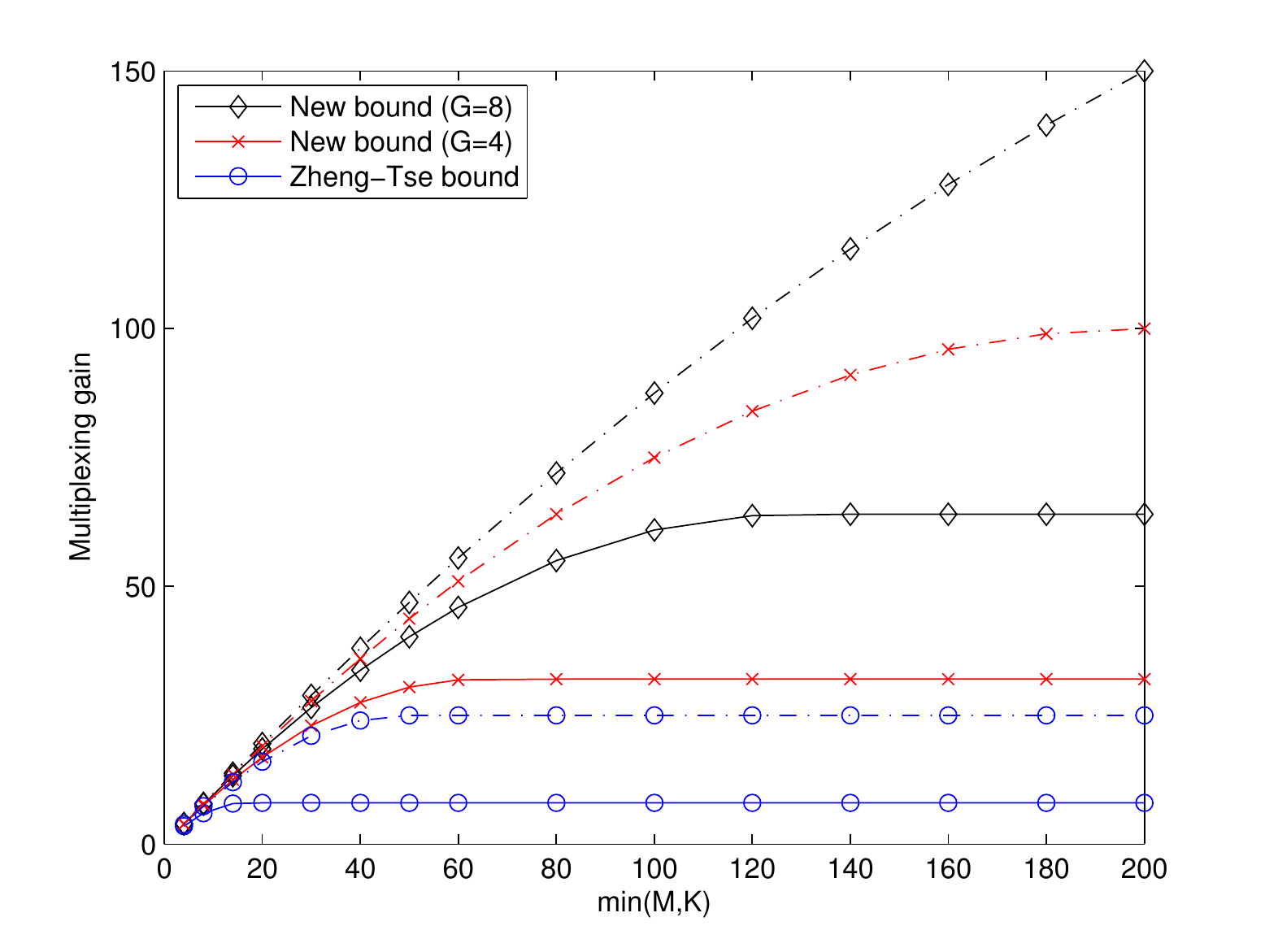} 
\caption{Impact of the degrees of transmit correlation diversity ($G$) on multiplexing gain for different numbers of $\min(M,K)$ in pilot-aided systems, where the solid lines indicate $T_c=32$ and the dash-dotted lines indicate $T_c=100$.
} \label{fig-1}
\end{figure}


\begin{ex}
Let $T_c$ take either 32 long-term evolution (LTE)  symbol duration \cite{LTE12} (approximately 60 km/h) or 100 symbol duration (19 km/h).  
Also, suppose that the unitary condition is attained such that $G=4$ and $G=8$. Fig. \ref{fig-1} shows the Zheng-Tse upper bound, $M_\text{iid}^*(1-M_\text{iid}^*/T_c)$, and the new bound, $M^*(1-M^*/T_cG)$, on the system multiplexing gain as $\min\{M,K\}$ increases. It can be seen that exploiting transmit correlation diversity can increase the multiplexing gain by a factor of $4$ for $G=4$ and $8$ for $G=8$, respectively.
\end{ex}

In the following, we consider two asymptotic cases. While the first case is when $G$ becomes large for $r$ fixed as $M$ increases, the other is for $G$ fixed with $r$ large. 

\subsubsection{Large $G$ Case}

So far, we have assumed a fixed number of degrees of transmit correlation diversity, $G$. We now turn our attention to the case where as $M\rightarrow \infty$, $G$ also grows such that the ratio $G/M$ is not vanishing (i.e., {bounded}). In practical systems, we generally consider a  large-scale array in the high carrier frequency $f_c$ due to the space limitation of large-scale array and the fact that the wavelength is inversely proportional to $f_c$. Then it is fairly reasonable to increase $M$ proportionally as $f_c$ grows and hence to let $M$ depend on $f_c$. It was observed, e.g., in mm-Wave channels \cite{Zha10}, that the higher  $f_c$, the smaller number of strong multipaths that the receivers experience due to higher directionality. This sparsity of dominant multipath components is also verified by mm-Wave propagation measurement campaigns \cite{Rap13}. Thus, the high transmit correlation diversity is attainable in the high $f_c$ case. Then, as both $M$ and $f_c$ grow, $G$ may continue increasing such that $G/M$ is fixed.

Since the impact of noisy CSIT on the system performance under the unitary structure was already addressed in \cite{Adh13}, we rather focus on the impact of pilot saving on the high-SNR capacity in this work. Thus, the BS can acquire noiseless (error-free) CSIT at the cost of either downlink common pilot or uplink dedicated pilot, according to (\ref{eq:PS-1}) and (\ref{eq:PS-2}).
Denote by $\mathcal{C}^\text{sum}_\text{p1}(P,M^*,\upsilon)$ the high-SNR capacity of pilot-aided system I for $M^*$ and $\upsilon$ large, where $\upsilon = G$ or $r$ and the subscript $\text{p1}$ indicates the pilot-aided system I. 
Assuming that the perfect CSIT is provided by an ideal (i.e., delay-free and error-free feedback) uplink with no channel estimation error in FDD and neither calibration error nor pilot contamination in TDD, respectively, $\mathcal{C}^\text{sum}_\text{p1}(P,M^*,\upsilon)$ is simply given by
\begin{align} 
  \mathcal{C}^\text{sum}_\text{p1}(P,M^*,\upsilon) &= \Big(1-\frac{q^*}{T_c}\Big)\; \Cc^\text{sum}(P,M^*,\upsilon). \nonumber 
\end{align}

In what follows, using the results of Section \ref{sec:IT}, we refine the $O(1)$ term in (\ref{eq:intro-4}) first in the large $G$ regime and then in the large $r$ regime. 
Let $$\mu_\text{p1}=\frac{M^*}{K}=\frac{q^*}{K'}$$ be fixed.
In this scenario, $G$ is taken to infinity along with $M$ but both $r$ and $K'$ are finite, unlike Theorem \ref{thm-8}. Therefore, we shall make use of  \cite[Thm. 2.11]{Tul04} instead of Lemma \ref{lem-2} then we can apply Theorem \ref{thm-2} in the sequel. 

\begin{thm} \label{thm-4}
Suppose the unitary structure and the uniform boundedness of $\lambda_{g,i}$ in (\ref{eq:SM-1c}). As $M\rightarrow \infty$, for $\mu< 1$,  the high-SNR capacity of the pilot-aided system I scales linearly in $M^*$ with the ratio
\begin{align} \label{eq:ID-15}
  \left(1-\frac{q^*}{T_c}\right)\log{\mu\epsilon} \;& +o(1) \le \frac{\mathcal{C}^\text{sum}_\text{p1}(P,M^*,G)}{M^*} - \left(1-\frac{q^*}{T_c}\right) \nonumber \\ 
  &\ \ \times \left\{{\log\frac{P}{q^*}}+\log e\Bigg(-\gamma+\sum_{\ell=2}^{K'}\frac{1}{\ell}+\Big(\frac{1-\mu_\text{p1}}{\mu_\text{p1}}\Big)\sum_{\ell=(1-\mu_\text{p1})K'+1}^{K'}\frac{1}{\ell}\Bigg)\right\} \le o(1).
\end{align}

For $\mu\ge 1$, the high-SNR capacity scales linearly in $M^*$ with the ratio
\begin{align}  \label{eq:ID-13b}
  \left(1-\frac{q^*}{T_c}\right)\log\epsilon \;&+o(1) \le \frac{\mathcal{C}^\text{sum}_\text{p1}(P,M^*,G)}{M^*} - \left(1-\frac{q^*}{T_c}\right) \left\{{\log\frac{P}{q^*}}+\log e\Bigg(-\gamma+\sum_{\ell=2}^{K'}\frac{1}{\ell}\Bigg)\right\} \le o(1).
\end{align}
\end{thm}

\begin{IEEEproof} 
For $\mu < 1$ and $q^*< T_c/2$, we have $M^*=M, q^*=r, \mu_\text{p1}=\mu$. In this case, we use (\ref{eq:IT-13}) instead of (\ref{eq:ID-5}). Then, the growth rate at which the high-SNR capacity increases in the large $G$ regime as $M\rightarrow \infty$ is lower-bounded by
\begin{align}  \label{eq:ID-13}
   \frac{\mathcal{C}^\text{sum}(P,M,G)}{M} &\ge  \log\frac{P}{M}+\log e\sum_{g=1}^G \frac{r}{M} \Bigg(-\gamma +\sum_{\ell=2}^{K'}\frac{1}{\ell}+\frac{K'-r}{r}\sum_{\ell=K'-r+1}^{K'}\frac{1}{\ell} \Bigg)\nonumber \\
   &\ \ \ \ \ +\frac{1}{M}\sum_{g=1}^G\log|\Lambdam_g|+o(1) \nonumber \\
   &= {\log\frac{P}{r}}+\log e\Bigg(-\gamma+\sum_{\ell=2}^{K'}\frac{1}{\ell}+\Big(\frac{1-\mu}{\mu}\Big)\sum_{\ell=(1-\mu)K'+1}^{K'}\frac{1}{\ell} \Bigg)+\log{\mu\epsilon}+o(1) 
\end{align}
where we used (\ref{eq:SM-1c}) since $\frac{\lambda_\text{min}}{G}\ge \frac{\lambda_\text{min}}{\lambda_\text{max}}$. We can similarly get the upper bound in (\ref{eq:ID-15}) for $q^*< T_c/2$. 
When $q^*={T_c}/{2}$, the rate of growth for the $\mu < 1$ case can be obtained in a similar way by noticing $M^*=\frac{T_cG}{2}$.
Therefore, for these two cases in the large $G$ regime with $q^*$ fixed, we get (\ref{eq:ID-15}).

For $q^*< T_c/2$ and $\mu \ge 1$, noticing that $M^*=K, q^*=K', \mu_\text{p1}=1$ and using (\ref{eq:IT-14}), 
we get 
\begin{align}  \label{eq:ID-14}
  \frac{\mathcal{C}^\text{sum}(P,K,G)}{K}   &\ge  \log\frac{P}{K}+\log e\sum_{g=1}^G \frac{K'}{K}\Bigg(-\gamma +\sum_{\ell=2}^{K'}\frac{1}{\ell}\Bigg)+\frac{1}{K}\sum_{g=1}^G \sum_{i=1}^{K'}\log \lambda_{g,i}+o(1) \nonumber \\
  &=  {\log\frac{P}{K'}}+\log e\Bigg(-\gamma+\sum_{\ell=2}^{K'}\frac{1}{\ell}\Bigg) +\log \epsilon+o(1)
\end{align}  
When ${q^*} = T_c/2$ and $\mu \ge 1$, the growth rate can be  obtained again by noticing $M^*=\frac{T_cG}{2}$. Then, we obtain (\ref{eq:ID-13b}) for $\mu \ge 1$.
\end{IEEEproof}

The following result shows the \emph{most optimistic} gain of transmit correlation diversity in the limit of $\Delta_g\rightarrow 0$ for all $g$, which we provide as a capacity upper bound even though this channel assumption may seem unrealistic. 

\begin{corol} \label{cor-4}
For $\mu=1$ and $T_c \ge 2$, as $\Delta_g\rightarrow 0$ and $M\rightarrow \infty$ with $\mu$  fixed, the high-SNR capacity of the pilot-aided system I scales linearly in $M$ with the ratio  
\begin{align} \label{eq:ID-1}
  \lim_{M \rightarrow \infty}\limsup_{\Delta_g\rightarrow 0}\frac{\mathcal{C}^\text{sum}_\text{p1}(P,M,G)}{M} = \left(1-T_c^{-1}\right) \log\frac{P}{e}. 
\end{align}
\end{corol}

To prove this, we first notice that the condition of (\ref{eq:intro-4}) can be restated as $T_c \ge 2 \min(r,K')$ in this case. Hence the sufficient condition is guaranteed just for $T_c \ge 2$, since $r\rightarrow 1$ as $\Delta_g\rightarrow 0$. Using this and (\ref{eq:ID-15}), (\ref{eq:ID-1}) immediately follows with ${r}/{T_c}=T_c^{-1}$.

It is remarkable that if the unitary structure is attained with $T_c$ sufficiently large and $\Delta_g$ sufficiently small, the high-SNR capacity of MU-MIMO systems approaches the full-CSI capacity in (\ref{eq:ID-23}).  The systems of interest are \emph{scalable} in $\min(M,K)$ and also the user throughput does not vanish any longer unless $M\ll K$ (i.e., $\mu\ll 1$), in sharp contrast to (\ref{eq:intro-1}). It should be pointed out that the growth rate in Corollary \ref{cor-4} was not obtained  just by pilot saving but also by the power gain due to eigen-beamforming. 

\subsubsection{Large $r$ Case}

In the large $r$ regime where $r$ goes to infinity while $G$ fixed, we can obtain the following result by using Theorem \ref{thm-8}.
For $\mu_\text{p1}\le 1$ 
\begin{align}  \label{eq:ID-16}
  \left(1-\frac{q^*}{T_c}\right)\log\mu_\text{p1}\epsilon +o(1) \le \frac{\mathcal{C}^\text{sum}_\text{p1}(P,M^*,r)}{M^*} -  \left(1-\frac{q^*}{T_c}\right)\left\{ \log \frac{P}{e\mu_\text{p1}}+\Big(\frac{1-\mu_\text{p1}}{\mu_\text{p1}}\Big)\log\frac{1}{1-\mu_\text{p1}}\right\} \le o(1).
\end{align}
Note that the $\mu_\text{p1}>1$ case does not happen in pilot-aided system I, since we make use of only $K'$ eigenmodes regardless of how large $r$ is, i.e., $q^*=K'$ according to (\ref{eq-0}). This restriction in system I may cause a nontrivial rate loss for $M> K$, as will be discussed in the next subsection.

\begin{figure} 
\center \includegraphics[scale=.65]{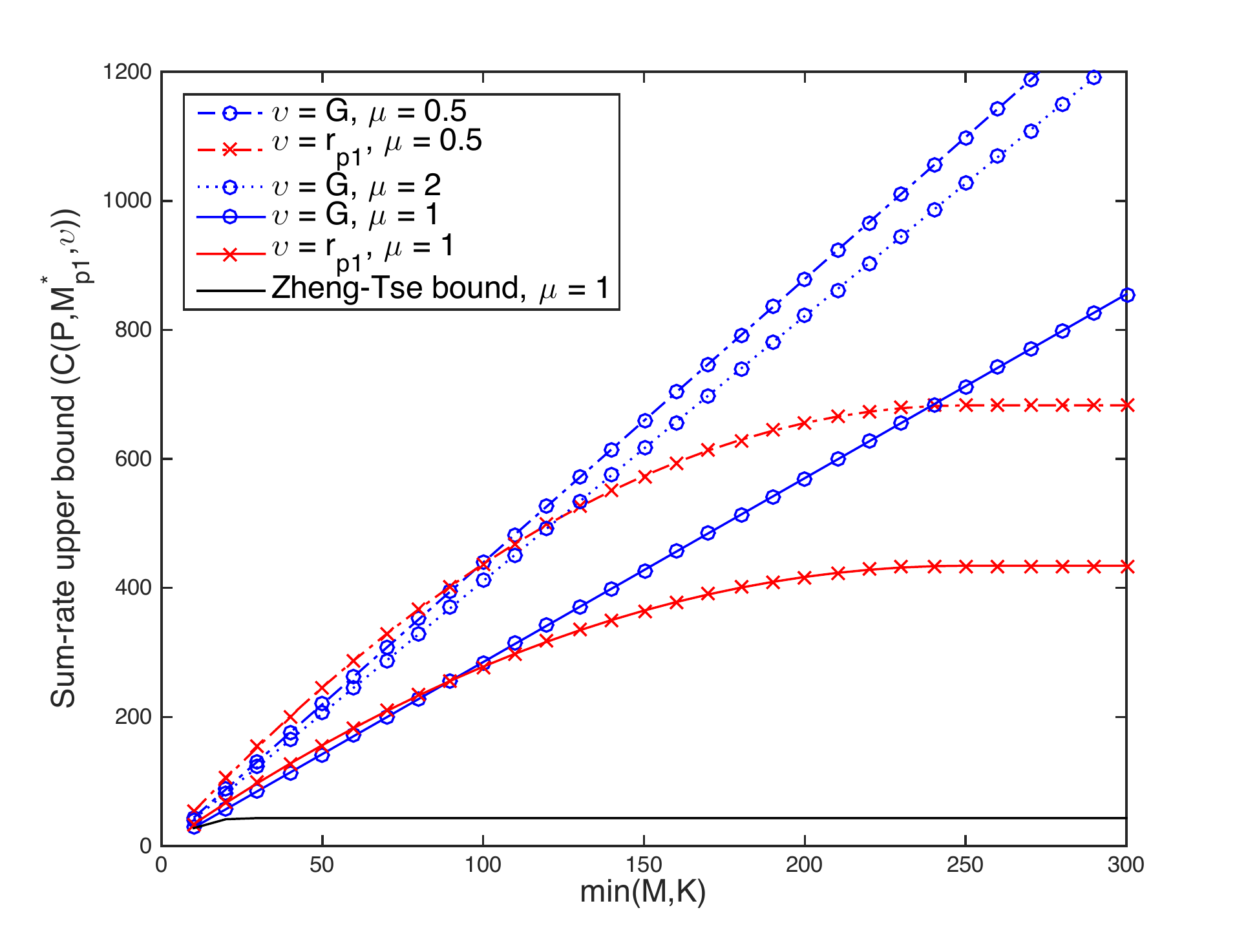} 
\caption{Asymptotic sum-rate upper bound curves versus $\min(M,K)$ in pilot-aided system I at $P=30$ with $T_c=50$, where $q^*=10$ when $G$ is large ($\upsilon=G$), and  $G=10$ when $q^*$ is large ($\upsilon=q^*$).} \label{fig-2}
\end{figure}

Fig. \ref{fig-2} shows the sum-rate upper bounds in (\ref{eq:ID-15}), (\ref{eq:ID-13b}), and (\ref{eq:ID-16}) on the asymptotic capacity for different system parameters in pilot-aided systems I. For large $M$, the  Zheng-Tse bound is given by $M_\text{iid}^*(1-M_\text{iid}^*/T_c)\log \frac{P}{e}+o(1)$.
For large $G$ and fixed $q^*$, the system multiplexing gain grows linearly with $\min\{M,K\}$, whereas this is not the case with large $q^*$ and fixed $G$. 
To understand the large rate gap between $\mu=1$ and $\mu=2$, recall the eigen-beamforming gain of up to $\log \mu G$ in Section \ref{sec:IT-D} and that a dual MAC is equivalent at high SNR to the corresponding MIMO point-to-point channel with $K$ transmit antennas and $M$ receive antennas. The equivalent MIMO channel is well understood to have a logarithmic power gain scaling with $M$ due to receive beamforming. For $\mu=0.5$ case, the large rate gap from $\mu=1$ is because the upper bound was given by allowing the intra-group cooperation within each group. 
Finally, for large $q^*$ but fixed $G$, the two cases of $\mu=1$ and $\mu=2$ collapse into the red solid line. This is due to the fact that pilot-aided system I considers only multiplexing gain but not power gain, which will be addressed in the following subsection.

\subsection{Pilot-Aided System II}

In the large $M$ regime, the $M>K$ ($\mu > 1$) case may be more frequently encountered in realistic systems, which is also the typical scenario of large-scale MIMO.  
We introduce a new pilot-aided system to address the foregoing issue for this case with $r$ large but $G$ fixed. In contrast to pilot-aided system I, in which only $K$ eigenmodes are used by letting $M^*=K$ when $K=\min\left\{M,K,\lfloor \frac{T_cG}{2}\rfloor\right\}$, we shall allow in the new system referred to as \emph{pilot-aided system II} to use more than $K$ eigenmodes, even though the degrees of freedom is certainly at most $K$. By doing so, we may obtain a noticeable power gain suggested by (\ref{eq:ID-6}) owing to transmit correlation diversity which compensates for the increase in channel uses required for downlink training. To understand this, notice that using more than $K$ eigenmodes has a smaller impact on the system multiplexing gain as $G$ and/or $T_c$ grows, as shown in (\ref{eq:intro-4}).

\begin{figure} 
\center\includegraphics[scale=.65]{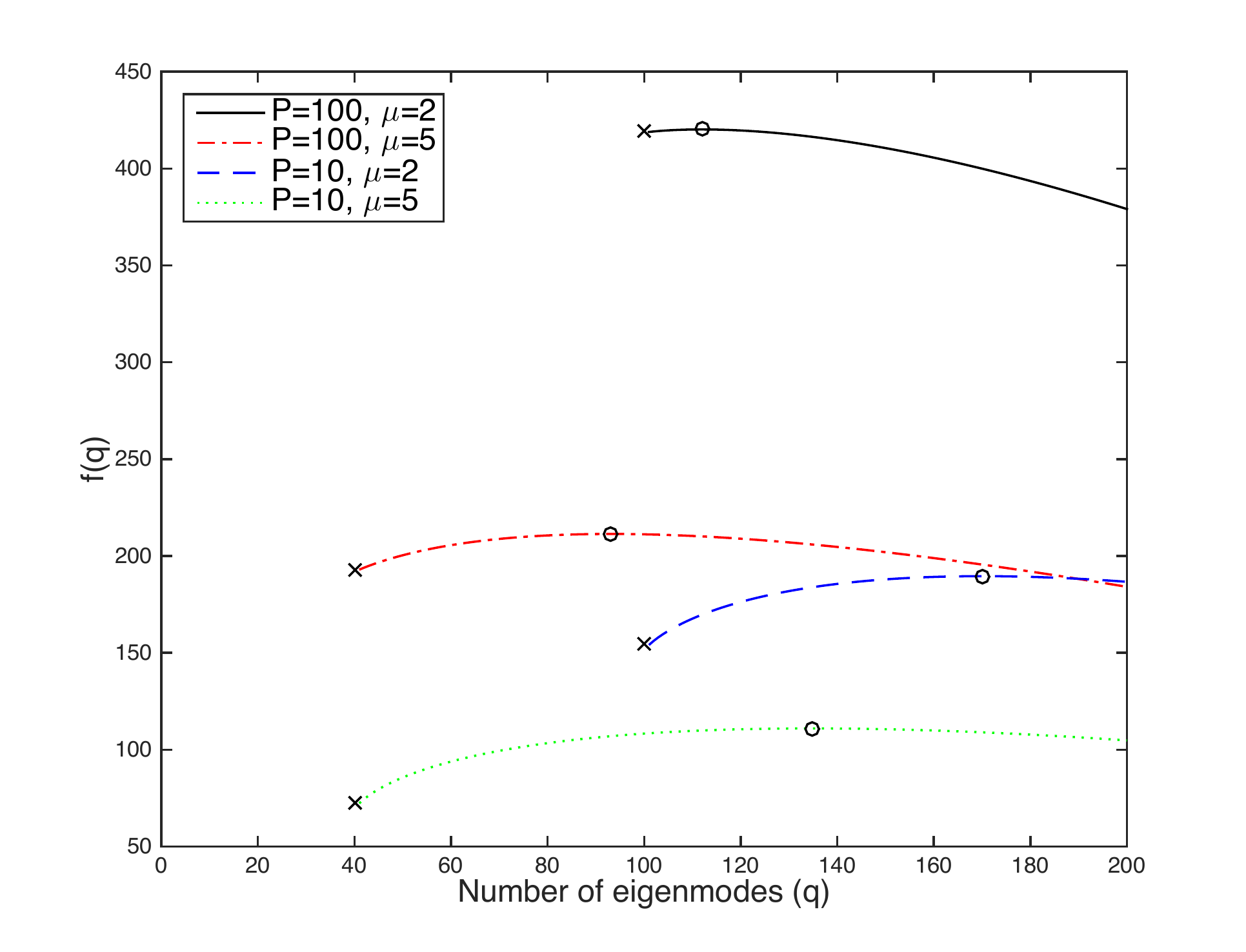} 
\caption{Values of $f(Q)$ versus the number of eigenmodes to use, $q$, in pilot-aided system II when $M> K$ (i.e., $\mu > 1$), where $M=200$, $T_c=64$, and the `o' indicates the optimum numbers of  eigenmodes, $M_\text{p2}^*$, and the `x' indicates $M^*$.} \label{fig-3}
\end{figure} 

To take into account the additional power gain from using more than $M^*$ eigenmodes in pilot-aided system II, we replace the optimization problem in (\ref{eq-01}) with the following one based on the upper bound in (\ref{eq:ID-6}).  
\begin{align} \label{ID-18}
   M_\text{p2}^* =\argmax_{q} f(q)
\end{align}
subject to $q^*\le q\le r$, where $f(q) = M^*\big\{(1-\frac{q}{T_c})\log \frac{P}{e}\frac{q}{K'}+(\frac{q}{K'}-1)\log\frac{q}{q-K'}\big\}$ with $M^*$ (the maximum number of degrees of freedom for the communication phase) unchanged and the subscript $\text{p2}$ indicates the pilot-aided system II. The high-SNR capacity of this new pilot-aided system for $\mu>1$ scales linearly in $K$ with the ratio
\begin{align}  \label{eq:ID-17}
 \Big(1-\frac{M_\text{p2}^*}{T_cG}\Big) \log \frac{\lambda_\text{min}}{G}&+o(1) \le \frac{\mathcal{C}^\text{sum}_\text{p2}(P,M_\text{p2}^*,r)}{K} \nonumber \\ &-\Big(1-\frac{M_\text{p2}^*}{T_cG}\Big)\left\{ \log \frac{\mu_\text{p2}P}{e}+(\mu_\text{p2}-1)\log\frac{\mu_\text{p2}}{\mu_\text{p2}-1}+c_{\text{p2},2}\right\} \le o(1)
\end{align}
where $\mu_\text{p2}=\frac{M_\text{p2}^*}{K}$.

\begin{figure} 
\center\includegraphics[scale=.65]{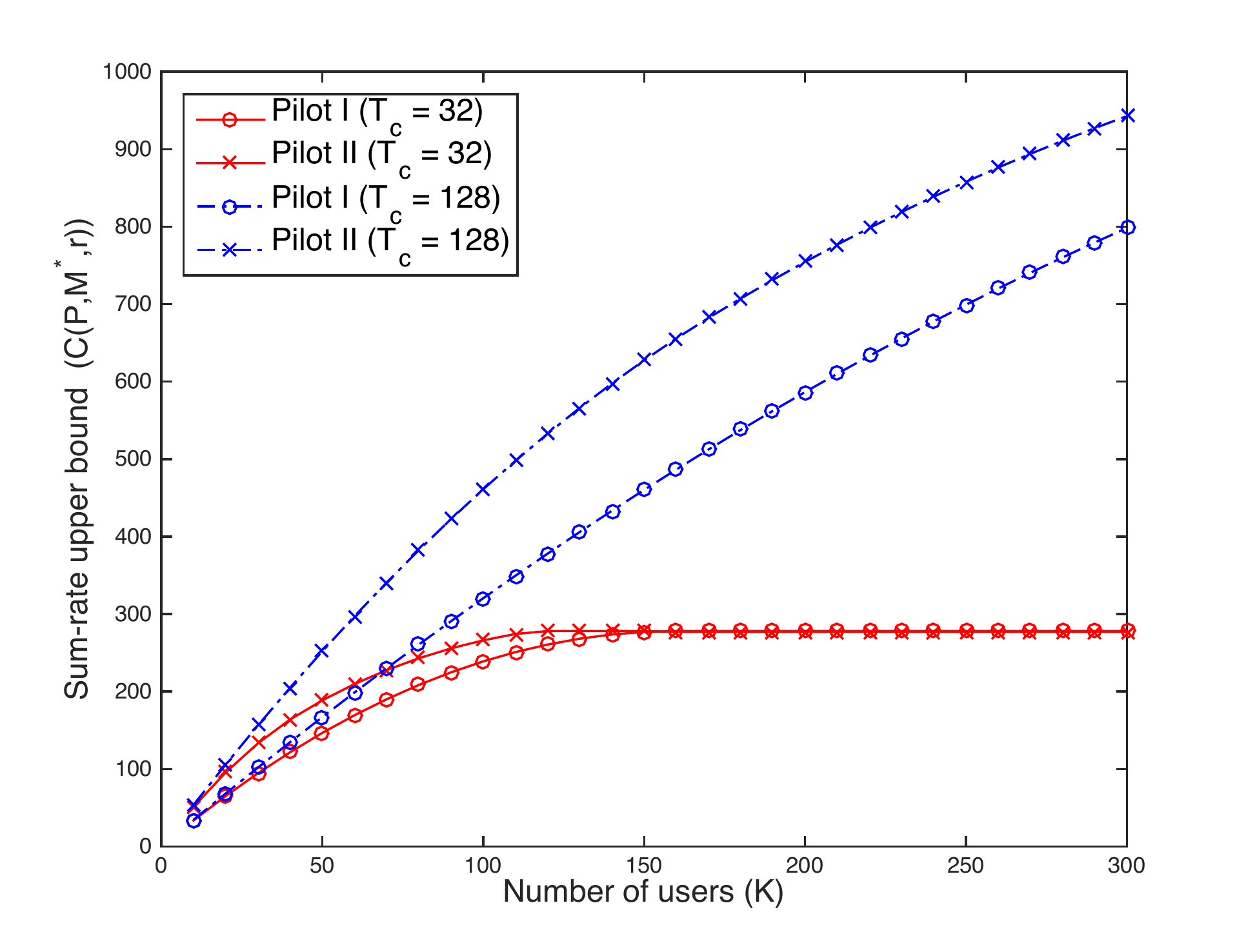} 
\caption{Asymptotic sum-rate upper bounds of two pilot-aided systems where $\mu=2$, $G=10$, and $P= 30$.} \label{fig-5}
\end{figure}

Fig. \ref{fig-3} shows the optimum number of eigenmodes, $M_\text{p2}^*$, for different $P$ and $\mu$ with $M=200$. Here, $M^*=K=40$ for $\mu=5$ and $M^*=K=100$ for $\mu=2$. Therefore, if we consider the power gain due to eigen-beamforming as well as the system multiplexing gain, the optimum values of $M_\text{p2}^*$ are shown to be quite different from $M^*$. We can also see that the resulting rate gap is reduced as $P$ increases for $T_c=64$.
Fig. \ref{fig-5} compares the asymptotic sum-rate upper bounds of pilot-aided system I and II when $T_c=32$ and $T_c=128$. It is shown that the rate gap gets larger as $T_c$ increases, since, for large $T_c$, the extra overhead due to training more than $K$ eigenmodes reduces, as mentioned earlier.

\begin{rem}
So far, we have assumed $T=1$ such that the channel covariances of all users associated to the BS satisfy a single unitary structure, which may seem far from practical systems. Recall that if we extend to the case of multiple classes as shown in Fig. \ref{fig-0}, each class should consume orthogonal time/frequency resources. Then the pilot design in Sec. \ref{PS-0} must increase the pilot overhead by a factor of $T$. Assuming homogeneous classes (the same number of homogeneous user groups per class), the pre-log factor in (\ref{eq:intro-4}) is then replaced with 
\begin{align} \label{eq-02}
   M^*\left(1-\frac{M^*T}{T_cG}\right)
\end{align}
where $M^*=\min\left\{M,K,\lfloor \frac{T_cG}{2T}\rfloor\right\}$.
This may undermine the potential gain of transmit correlation diversity. Therefore, we have the system design guideline that $T$ should be less than the number ($G$) of degrees of transmit correlation diversity and it must be restricted as small as possible. If $T\ge G$, there is no point in using the multiple pre-beamformed pilot in Sec. \ref{PS-0}. 
\end{rem}

Finally, it should be pointed out that transmit correlation diversity can promise a significant capacity gain in all regimes of interest, considering the cost of downlink training. Even though the new diversity may bring a power loss to the capacity compared to the independent fading case assuming perfect CSI in Sec. \ref{sec:IT}, the increase in multiplexing gain can fully offset the power loss unless $\min\{M,K\}$ is too small.

\section{Concluding Remarks}
\label{sec:CR}

In this paper, we have investigated several asymptotic capacity bounds of correlated fading MIMO BCs to understand the impact of transmit correlation on the capacity. In order to intuitively show and fully exploit the potential benefits of a new type of diversity --- transmit correlation diversity, we imposed the unitary structure on channel covariances of users. Assuming perfect CSIT without a cost of downlink training, we showed that transmit correlation diversity is not always beneficial to the high-SNR capacity of Gaussian MIMO BCs in all regimes of system parameters of interest such as $M,K,r,G$, and $T_c$. In particular, the new diversity is even detrimental to the capacity in the large-scale array regime.
Taking the cost for downlink training into account, however, we found that transmit correlation diversity is indeed very beneficial in all regimes of interest. Specifically, the system multiplexing gain can continue growing as the number of antennas and the number of users increase, as long as the degrees of transmit correlation diversity are sufficiently large. 
Even if we have focused on the downlink system in this work, notice that the new diversity can be leveraged in various forms of MIMO wireless networks including multi-cell uplink/downlink systems and wireless interference networks. 

It was shown that the eigen-beamforming gain due to pre-beamforming along large-scale  eigensapces of user groups is essential to achieve capacity benefits of transmit correlation diversity. This provides an insight that a precoding scheme which can realize a large portion of such a gain may be competitive to ZFBF with noisy or outdated CSIT for correlated fading channels particularly in the large $K$ regime. In order to validate this argument, our recent work in \cite{Nam14b} proposed a new limited feedback framework for large-scale MIMO systems. 

In MIMO wireless communications, there exist three most essential resources: time, frequency, and ``{small-scale}" space that depends on instantaneous channel realizations. Apart from these resources, we have identified the new type of resource, transmit correlation diversity (namely, ``{large-scale}" spatial resource), and provided some new insights on how to use it, when it is beneficial to the capacity, and how much it affects the system performance. The most remarkable result can be summarized as:
\emph{Exploiting transmit correlation may increase the multiplexing gain of the MU-MIMO system, when taking into account the downlink or uplink training overhead, 
by a factor equal to the degrees of transmit correlation diversity, i.e., the number of user groups with mutually orthogonal (or linearly independent for a weaker condition) channel covariance eigenspaces.}

\appendices

\section{Proof of Theorem \ref{thm-2}}  
\label{app:proof-2}

We first prove the case of $r<K'$. Provided the unitary structure is available, the sum rate of the $g$th  dual MAC subchannel in (\ref{eq:MU-0}) can be rewritten as
\begin{align} \label{eq:IT-40}
   \log \left | \Id + \Lambdam_g^{1/2} \Wm_g \Sm_g \Wm_g^\mathsf{H} \Lambdam_g^{1/2} \right | &=  \log \left | \Lambdam_g^{-1} +  \Wm_g \Sm_g \Wm_g^\mathsf{H}  \right | +\log \left | \Lambdam_g \right |.  
\end{align}
By allowing the intra-group cooperation (i.e., the receiver cooperation within each group) and following the standard argument, the capacity region of the dual MAC subchannel is outer-bounded by that of the corresponding cooperative MIMO system. Given the perfect CSIT and at high SNR, the asymptotic optimal input $\Xm_g$ in the cooperative MIMO system is the uniform power allocation over $r$ eigenmodes of $\pW_g\pW_g^\ct$  with $\sum_g\trace(\Xm_g)\le P$, since the Wishart matrix $\pW_g\pW_g^\ct$ is well conditioned with high probability for all $g$. Here, the difference with our problem of interest is that the noise variances at $r$ effective antennas of the receiver in the $g$th dual MAC in the RHS of (\ref{eq:IT-40}) are scaled by $\lambda_{g,i}$, where $i=1,\cdots,r$. But this does not change the known result. 
Then, we have at high SNR (i.e., large $P$)
\begin{align} \label{eq:IT-35}
   \mathbb{E} \Big[\max_{\trace(\pS_g)\le P}\log \left | \Lambdam_g^{-1} +  \Wm_g \Sm_g \Wm_g^\mathsf{H}  \right | \Big ]  &\le \mathbb{E} \Big[\max_{\trace(\pX_g)\le P}\log \left | \Lambdam_g^{-1} +  \Wm_g \Xm_g \Wm_g^\mathsf{H}  \right | \Big ] \nonumber \\
   &\simeq \mathbb{E} \bigg[\log \Big | \Lambdam_g^{-1} +  \frac{P}{M}\Wm_g \Wm_g^\mathsf{H}  \Big | \bigg ] \nonumber \\
   &\simeq  \mathbb{E} \Big[\log \big |   \Wm_g \Wm_g^\mathsf{H}  \big |\Big ] +r\log \frac{P}{M}
\end{align}
where $\simeq$ denotes the asymptotic equivalence (the difference between both sides vanishes as $P\rightarrow \infty$).
As a consequence, when $r<K'$, the sum capacity is upper-bounded as 
\begin{align} 
  \mathcal{C}^\text{sum}(P) \le & \ M\log\frac{P}{M}+\sum_{g=1}^G \mathbb{E} \Big [\log\left|  \Wm_g \Wm_g^\ct \right|\Big ] +\sum_{g=1}^G\log|\Lambdam_g|+o(1) \label{eq:IT-37a} \\
  = & \ M\log\frac{P}{M}+ rG\Bigg(-\gamma +\sum_{\ell=2}^{K'}\frac{1}{\ell}+\frac{K'-r}{r}\sum_{\ell=K'-r+1}^{K'}\frac{1}{\ell} \Bigg)\log e +\sum_{g=1}^G\log|\Lambdam_g|+o(1) \label{eq:IT-37}
\end{align}
where we used the well-known result of random matrix theory in \cite[Thm. 2.11]{Tul04}, namely,  $\mathbb{E} \Big [\ln\left|  \Wm_g \Wm_g^\ct \right|\Big ]  =  \sum_{\ell=0}^{r-1} \psi(K'-\ell)$, where $\Wm_g \Wm_g^\ct$ is almost surely nonsingular and $\psi(n) = -\gamma +\sum_{\ell=1}^{n-1}\frac{1}{\ell}$ with $\gamma$ the Euler-Mascheroni constant. From (\ref{eq:ID-11}), the second equality in (\ref{eq:IT-37}) immediately follows.

The lower bound in (\ref{eq:IT-13}) is given by simply letting the diagonal input matrix $\pS_g$ as $\pS_g=  \frac{P}{K} \pI_{K}$ for all $g$, which is in fact the optimal input covariance when only the channel distribution is accessible at the receiver in MIMO MAC with each user having the same power constraint \cite{Gol03}.  
The resulting $\Wm_g \Sm_g \Wm_g^\mathsf{H} = \frac{P}{K}{\Wm}_g {\Wm}_g^\mathsf{H}$ is also a Wishart matrix with $r$ degrees of freedom and hence  
\begin{align} \label{eq:IT-35b}
   \log \left | \Lambdam_g^{-1} +  \Wm_g \Sm_g \Wm_g^\mathsf{H}  \right |  &\ge  \log \left |  {\Wm}_g {\Wm}_g^\mathsf{H} \right | +r\log \frac{P}{K}+o(1)
\end{align}
Using \cite[Thm 2.11]{Tul04} again, we have
\begin{align}  
  \mathcal{C}^\text{sum}(P) \ge  & \ M\log\frac{P}{K}+ rG\Bigg(-\gamma +\sum_{\ell=2}^{K'}\frac{1}{\ell}+\frac{K'-r}{r}\sum_{\ell=K'-r+1}^{K'}\frac{1}{\ell} \Bigg)\log e  +\sum_{g=1}^G\log|\Lambdam_g|+o(1) .
\end{align}
Then, the high-SNR capacity upper and lower bounds differ by $M\log \frac{r}{K'}$.

Next, we consider the second case of $r \ge K'$. 
When the number of transmit antennas is greater than or equal to the total number of receive antennas in a MIMO BC, the sum capacity of its dual MAC is well known \cite{Cai03} to be equivalent at high SNR to that of the corresponding point-to-point MIMO system. This also implies that the uniform power allocation across $K'$ eigenmodes, i.e., $\pS_g=\frac{P}{K}\pI_{K'}$, is asymptotically optimal for the $g$th dual MAC (\ref{eq:MU-0}) equivalent to the the point-to-point channel where the number of receive antennas ($M$) is larger than the number of transmit antennas ($K$). Since transmit correlation is only harmful to the capacity of the equivalent point-to-point channel 
for perfect CSI and large $P$, the capacity of each dual MAC is upper-bounded by the i.i.d. Rayleigh fading channel where $\lambda_{g,i}=\frac{M}{r}=G$ for all $(g,i)$ due to $\trace(\Lambdam_g)=M$.
Then, we have 
\begin{align} \label{eq:IT-28b}
      \log \left | \Id_r + \Lambdam_g^{1/2} \Wm_g \Sm_g \Wm_g^\mathsf{H} \Lambdam_g^{1/2} \right |  
   &\simeq \log \left | \Id_r + \frac{P}{K}\Lambdam_g^{1/2} \Wm_g  \Wm_g^\mathsf{H} \Lambdam_g^{1/2} \right |  \nonumber \\
   &\le \log \left | \Id_{K'} + \frac{PG}{K} \Wm_g^\ct \Wm_g \right |   \nonumber \\
   &\simeq  \log \left |   \Wm_g^\ct \Wm_g \right | +K'\log \frac{P}{K} +K'\log G. 
\end{align}
For the lower bound, we can get
\begin{align} \label{eq:IT-45}
      \log \left | \Lambdam_g^{-1} +  \Wm_g \Sm_g \Wm_g^\mathsf{H}  \right |
   &\simeq \log \left | \Lambdam_g^{-1} +  \frac{P}{K}\Wm_g  \Wm_g^\mathsf{H}  \right |  \nonumber \\
   &\overset{(a)}{\ge} \log \prod_{i=1}^{r} \left ( \lambda_{g,r-i+1}^{-1} +  \frac{P}{K}\lambda_i(\Wm_g  \Wm_g^\mathsf{H}) \right )  \nonumber \\
   &\overset{(b)}{=} \log \prod_{i=1}^{K'} \left ( \lambda_{g,r-i+1}^{-1} +  \frac{P}{K}\lambda_i(\Wm_g^\ct \Wm_g) \right ) +\log \prod_{i=K'+1}^{r} \lambda_{g,r-i+1}^{-1}  \nonumber \\
   &\simeq  \log \left |   \Wm_g^\ct \Wm_g \right | +K'\log \frac{P}{K} +\log \prod_{i=K'+1}^{r} \lambda_{g,r-i+1}^{-1} 
\end{align}
where $(a)$ follows from the following lower bound on the determinant of the sum of two Hermitian matrices\cite{Fie71}: 
Let $\Am$ and $\Bm$ be Hermitian matrices with eigenvalues $\lambda_1(\Am)\ge\lambda_2(\Am)\ge \cdots \ge \lambda_n(\Am)$ and $\lambda_1(\Bm)\ge\lambda_2(\Bm)\ge \cdots \ge \lambda_n(\Bm)$, respectively. If $\lambda_n(\Am)+\lambda_n(\Bm)\ge 0$, then
\begin{align} \label{UP-14}
  \prod_{i=1}^{n} \left(\lambda_i(\Am)+\lambda_{i} (\Bm) \right ) \le \big|\Am+\Bm\big| \le   \prod_{i=1}^{n} \left(\lambda_i(\Am)+\lambda_{n-i+1} (\Bm) \right ) .
  \end{align}
In $(b)$, we used from the fact that the non-zero eigenvalues of $\Wm_g  \Wm_g^\mathsf{H}$ are the same as those of $\Wm_g^\ct \Wm_g$. Similarly using the upper bound in (\ref{UP-14}), we can obtain another high-SNR upper bound based on the same inequalities in \cite{Fie71} as 
\begin{align} 
      \log \left | \Lambdam_g^{-1} +  \Wm_g \Sm_g \Wm_g^\mathsf{H}  \right | + \log|\Lambdam_g|
   &\le  \log \left |   \Wm_g^\ct \Wm_g \right | +K'\log \frac{P}{K} +\log \prod_{i=1}^{K'} \lambda_{g,i}. \nonumber
\end{align}
However, the upper bound in (\ref{eq:IT-28b}) is clearly tighter than this one.

Using (\ref{eq:IT-28b}), (\ref{eq:IT-45}), and the fact that, for $r \ge K'$, the Wishart matrix $\Wm_g^\ct \Wm_g$ is almost surely nonsingular and hence invoking \cite[Thm 2.11]{Tul04} again, we have 
\begin{align}  
  \sum_{g=1}^G&\log \prod_{i=1}^{K'} \lambda_{g,r-i+1}+o(1) \nonumber \\
 & \le   \mathcal{C}^\text{sum}(P) -K \log\frac{P}{K} +K\Bigg(-\gamma +\sum_{\ell=2}^{r}\frac{1}{\ell} +\frac{r-K'}{K'}\sum_{\ell=r-K'+1}^{r}\frac{1}{\ell}\Bigg)\log e    \nonumber \\ &\le K\log G+o(1).
\end{align}
Therefore, we have (\ref{eq:IT-14}). This completes the proof.

\section{Achievability of (\ref{eq:TC-3})}  
\label{app:proof-3}

The achievability proof of (\ref{eq:TC-3}) begins with (\ref{eq:IT-40}), Corollary 1 in \cite{Nam13a}, and the uniform power allocation over groups such that $\pS_g=\frac{P}{M}\pI_{r}$, yielding
\begin{align}  \label{eq:TC-3b}
  \sum_{g=1}^G r\log\log K' +M\log\frac{P}{M}+\sum_{g=1}^G\log|\Lambdam_{g}|+o(1).
\end{align}
Compared to $M\log\log K$ in the i.i.d. Rayleigh fading case, the multiuser diversity gain reduces to $\sum_{g=1}^G r\log\log K'$. To show that the loss due to this diversity gain reduction (or the channel dimension reduction) vanishes for sufficiently large $K'$, we use the logarithmic identity
\begin{align} \label{eq:IT-31}
  \log_c(a\pm b)=\log_c a+\log_c\left(1\pm\frac{b}{a}\right)
\end{align}
where $a$ and $b$ are nonnegative. Then, we get 
$$\sum_{g=1}^G r\log\log K'=M\log\log K+o(1)$$
for large $K'$. This proves the achievability.

\section{Proof of Theorem \ref{thm-8}}  
\label{app:proof-8}

The proof begins with the dual MAC in  (\ref{eq:MU-0}) divided by $M$
\begin{align} \label{eq:ID-7}
   \frac{1}{M}\sum_{g=1}^G&\log \left | \Id + \Lambdam_g^{1/2} \Wm_g \Sm_g \Wm_g^\mathsf{H} \Lambdam_g^{1/2} \right | \nonumber \\
   &= \frac{1}{M} \sum_{g=1}^G\log \left | \Lambdam_g^{-1} +  \Wm_g \Sm_g \Wm_g^\mathsf{H}  \right | +\frac{1}{M}\sum_{g=1}^G \log \left | \Lambdam_g \right |
\end{align}
where the equality is given by (\ref{eq:IT-40}) and the assumptions. 

The following lemma shows a useful asymptotic behavior of the central Wishart matrix.
 
\begin{lem} \label{lem-2}
For $m$ large with the ratio $\eta=\frac{n}{m}$ fixed, the central Wishart matrix $\Wm\Wm^\ct$ with $\Wm$ the $m\times n$ matrix, where $n\ge m$, shows the asymptotic behavior
\begin{align} \label{eq:ID-4}
   \frac{1}{m}\mathbb{E} \left [\ln\left|  \Wm\Wm^\ct \right|\right ]  =  \left(\eta-1\right)\ln \frac{\eta}{\eta-1} +\ln n-1 +O(m^{-1}).
\end{align}
\end{lem}

\begin{IEEEproof}
The proof of (\ref{eq:ID-4}) can be immediately given by applying 
\begin{align} \label{eq:ID-11}
  \frac{1}{k}\sum_{\ell=1}^{k} \psi(\ell)=\psi(k+1)-1
\end{align} 
and by using the fact that $\psi(k)$ behaves as
\begin{align} \label{eq:ID-10}
   \lim_{k\rightarrow \infty}\psi(k) =\ln k +O(k^{-1})
\end{align} 
due to $\lim_{k\rightarrow \infty} \sum_{n=1}^{k} \frac{1}{n} - \ln k = \gamma$.
\end{IEEEproof}

For $\mu < 1$ (i.e., $r<K'$) at high SNR ($P$), taking expectation on the first term in the right-hand side (RHS) of (\ref{eq:ID-7}), we have the upper bound
\begin{align} \label{eq:ID-8}
   \frac{1}{M}\mathbb{E}&\left [\sum_{g=1}^G\log \left | \Lambdam_g^{-1} +  \Wm_g \Sm_g \Wm_g^\mathsf{H}  \right | \right ] \nonumber \\
   &\overset{(a)}{\le}  \frac{1}{r}\;\mathbb{E} \Big [\log \left |   \Wm_g \Wm_g^\mathsf{H}  \right | \Big ] +\log \frac{P}{M}  \nonumber \\
   &\overset{(b)}{=} \log e \left\{(\mu^{-1}-1)\ln\frac{\mu^{-1}}{\mu^{-1}-1} +\ln K'-1+O(r^{-1}) \right\} +\log \frac{P}{M}\nonumber \\
   &= \log \frac{P}{e\mu G}+\Big(\frac{1-\mu}{\mu}\Big)\log\frac{1}{1-\mu} +O(r^{-1})
\end{align}
where $(a)$ follows from (\ref{eq:IT-35}) and $(b)$ follows from (\ref{eq:ID-4}) in Lemma \ref{lem-2}. From (\ref{eq:IT-35b}) and (\ref{eq:ID-4}), we also get the lower bound
\begin{align} \label{eq:ID-9}
   \frac{1}{M}\mathbb{E}\left [\sum_{g=1}^G\log \left | \Lambdam_g^{-1} +  \Wm_g \Sm_g \Wm_g^\mathsf{H}  \right | \right ] 
   &\ge \log \frac{P}{e G}+\Big(\frac{1-\mu}{\mu}\Big)\log\frac{1}{1-\mu} +O(r^{-1}).
\end{align}

The second term in the RHS of (\ref{eq:ID-7}) can be bounded by using 
\begin{align} \label{eq:ID-20}
   \log\lambda_\text{min} \le \frac{1}{M}\sum_{g=1}^G\log \left | \Lambdam_g \right | \le \frac{1}{M}\sum_{g=1}^G\log \left(\frac{\trace(\Lambdam_g)}{r}\right)^r =\log G
\end{align}
where we used $\trace(\pR_g)=\trace(\Lambdam_g)=M$ and the geometric-arithmetic mean inequality $|\pA| \le \big(\frac{\trace(\pA)}{a}\big)^a$ with $a$ being the rank of an $n\times n$ matrix $\pA$. 
Combining (\ref{eq:ID-8}) -- (\ref{eq:ID-20}), we obtain (\ref{eq:ID-5}).  

For $\mu \ge 1$ (i.e., $r\ge K'$) and large $K'$, similar to the above steps with 
\begin{align}
    \log\lambda_\text{min} &\le \frac{1}{K} \sum_{g=1}^G\log\prod_{i=1}^{K'} \lambda_{g,r-i+1} \nonumber \\ &\le \frac{1}{K} \sum_{g=1}^G\log\prod_{i=1}^{K'} \lambda_{g,i}\nonumber \\ 
    &\le \log  G  \label{eq:ID-21}
\end{align}
where the last inequality follows from (\ref{eq:IT-28b}), we can obtain (\ref{eq:ID-6}) by using (\ref{eq:IT-28b}), (\ref{eq:IT-45}) and (\ref{eq:ID-4}). 
The remaining details are omitted for the sake of the compactness of this paper.

\bibliographystyle{IEEEtran}
\bibliography{transmit_correlation_diversity}

\end{document}